\documentclass[preprint2]{aastex631}
\usepackage{showyourwork}
\usepackage{macros}
\usepackage{amsmath}
\usepackage{enumerate}
\usepackage{acro}
\acsetup{patch/longtable=false}

\DeclareAcronym{LVK}{short=LVK, long=LIGO-Virgo-KAGRA}
\DeclareAcronym{O5}{short=O5, long=the fifth LIGO-Virgo-KAGRA observing run}

\DeclareAcronym{GW}{short=GW, long=gravitational wave}
\DeclareAcronym{CBC}{short=CBC, long=compact binary coalescence}
\DeclareAcronym{BBH}{short=BBH, long=binary black hole}

\DeclareAcronym{GP}{short=GP, long=Gaussian process, long-plural=es}

\begin{document}

\title{No need to know: towards astrophysics-free gravitational-wave cosmology}


\author[0000-0002-6121-0285]{Amanda M. Farah}
\email{afarah@uchicago.edu}
\affiliation{Department of Physics, University of Chicago, Chicago, IL 60637, USA}

\author[0000-0001-9892-177X]{Thomas A. Callister}
\affiliation{Kavli Institute for Cosmological Physics, The University of Chicago, Chicago, IL 60637, USA}

\author[0000-0002-7213-3211]{Jose Mar\'ia Ezquiaga}
\affiliation{Niels Bohr International Academy, Niels Bohr Institute, Blegdamsvej 17, DK-2100 Copenhagen, Denmark}

\author[0000-0002-0147-0835]{Michael Zevin}
\affiliation{The Adler Planetarium, 1300 South DuSable Lake Shore Drive, Chicago, 60605, IL, USA}

\author[0000-0002-0175-5064]{Daniel E. Holz}
\affiliation{Department of Physics, University of Chicago, Chicago, IL 60637, USA}
\affiliation{Kavli Institute for Cosmological Physics, The University of Chicago, Chicago, IL 60637, USA}
\affiliation{Department of Astronomy \& Astrophysics, The University of Chicago, Chicago, IL 60637, USA}
\affiliation{Enrico Fermi Institute, The University of Chicago, Chicago, IL 60637, USA}

\begin{abstract}
\Acp{GW} from merging compact objects encode direct information about the luminosity distance to the binary. 
When paired with a redshift measurement, this enables standard-siren cosmology: a Hubble diagram can be constructed to directly probe the Universe's expansion.
This can be done in the absence of electromagnetic measurements, as features in the mass distribution of \ac{GW} sources provide self-calibrating redshift measurements without the need for a definite or probabilistic host galaxy association. 
This ``spectral siren'' technique has thus far only been applied with simple parametric representations of the mass distribution, and theoretical predictions for features in the mass distribution are commonly presumed to be fundamental to the measurement. 
However, the use of an inaccurate representation leads to biases in the cosmological inference, an acute problem given the current uncertainties in true source population.
Furthermore, it is commonly presumed that the form of the mass distribution must be known \emph{a priori}\/ to obtain unbiased measurements of cosmological parameters in this fashion.
Here, we demonstrate that spectral sirens can accurately infer cosmological parameters without such prior assumptions.
We apply a flexible, non-parametric model for the mass distribution of compact binaries to a simulated catalog of 1,000 \ac{GW} signals, consistent with expectations for the next \acl{LVK} observing run.
We find that, despite our model's flexibility, both the source mass model and cosmological parameters are correctly reconstructed.
We predict a $%
  11.2
\unskip\label{output/nonparh0percent.txt}\unskip%
\%$ measurement of \Ho{}, keeping all other cosmological parameters fixed, and a $%
  6.4
\unskip\label{output/Hz_percent.txt}\unskip%
\%$  measurement of $H(z=%
  0.9
\unskip\label{output/mostsensitivez.txt}\unskip%
)$ when fitting for multiple cosmological parameters ($1\sigma$ uncertainties).
This astrophysically-agnostic spectral siren technique will be essential to arrive at precise and unbiased cosmological constraints from \ac{GW} source populations.
\end{abstract}

\section{Introduction}
\label{sec:intro}
Like light, \acp{GW} are redshifted as they propagate across the universe, thereby bearing imprints of the Universe's cosmic expansion history.
Unlike light, however, the form of \ac{GW} signals are known from first principles, directly from the theory of general relativity. Furthermore, because \acp{GW} propagate across the Universe without attenuation from intervening matter, and because the properties of \ac{GW} detectors are well characterized, \ac{GW} selection effects are extremely well understood. 
This allows for a precise estimate of each \ac{GW} catalogs' completeness and an unbiased measurement of the true \ac{GW} source population~\citep{2023PhRvX..13d1039A,abbott_population_2023,2023PhRvD.108d3011E}. 
Additionally, the \ac{GW} signals observed by the LIGO, Virgo, and KAGRA detectors~\citep{aasi_advanced_2015,acernese_advanced_2014,akutsu_overview_2021}  provide direct measurements of the distance to their sources.
This makes them ``standard sirens'': direct probes of cosmological parameters that circumvent the need for a cosmological distance ladder~\citep{schutz_determining_1986,holz_using_2005}. 

A well-known demonstration of standard siren cosmology was the multi-messenger event GW170817 \citep{abbott_multi-messenger_2017,coulter_swope_2017,valenti_discovery_2017,2017ApJ...848L..27T}, whose clear association with a host galaxy provided a precise redshift measurement and allowed for a direct ``bright siren'' measurement of the Hubble constant, \Ho{} \citep{abbott_gravitational-wave_2017}.
External redshift information can also come from galaxy catalogs, which provide an ensemble of possible redshifts for each \ac{GW} signal, allowing for a probabilistic ``dark siren'' measurement of \Ho{} when multiple \ac{GW} detections are combined \citep{del_pozzo_inference_2012, chen_two_2018, fishbach_standard_2019, soares-santos_first_2019,gray_cosmological_2020, abbott_gravitational-wave_2021, gwtc3_cosmo, gray_pixelated_2022,gray_joint_2023,mastrogiovanni_joint_2023, gair_hitchhikers_2023}.

Electromagnetic information about \ac{GW} source redshifts need not be available in order to use them as standard sirens, however.
\ac{GW} signals provide direct measurements of each source's luminosity distance, $D_L$, and redshifted (detector frame) masses, $m_{\det}= m_{\source}(1+z)$ \citep[e.g.][]{chen_mass-redshift_2019}.
Therefore, if the source frame mass is known, each \ac{GW} signal provides a direct mapping between luminosity distance and redshift, allowing for a measurement of the expansion of the universe at the time the \ac{GW} signal was emitted, $H(z)$.
\begin{figure*}
    \centering
    \includegraphics[width=\textwidth]{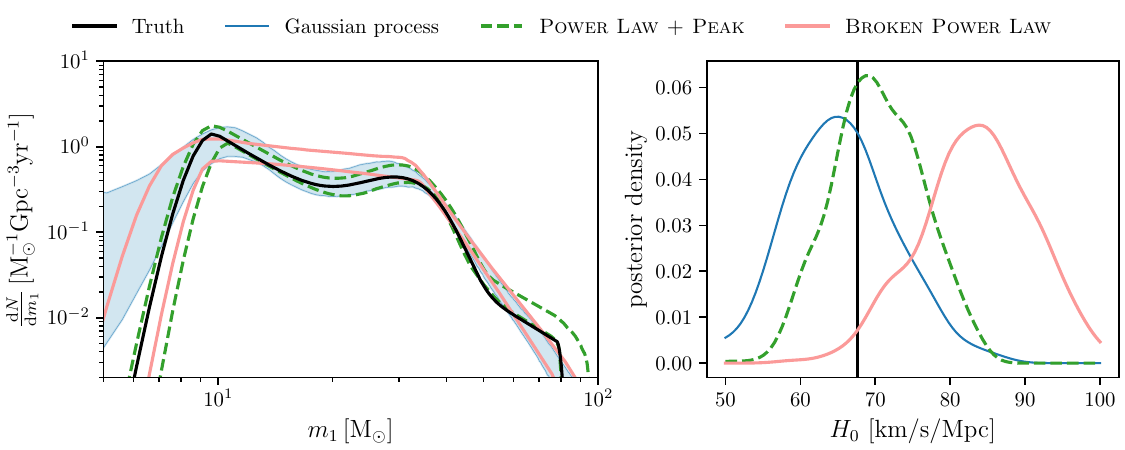}
    \caption{Spectral siren measurement for a simulated catalog with the correct parametric model (\plp, green dashed), a deliberately incorrect parametric model (\textsc{Broken Power Law}, pink solid) and the non-parametric model presented in this work (\acl{GP}, blue filled).
    The left panel shows the recovered source frame primary mass distribution for each model, and the simultaneously-inferred posteriors on \Ho{} are shown in the right panel.
    The bands in the left panel represent the 90\% credible intervals.
    The mass distribution and \Ho{} value used to generate the data are shown by a solid black line in each panel. 
    The deliberately incorrect parametric model fails to recover the true mass distribution and therefore produces an estimate of \Ho{} that is offset from the true value, whereas both the correct and non-parametric models recover the mass distribution and \Ho{}.
    As the true mass distribution is unknown for real observations, using a non-parametric model mitigates systematic uncertainty that would otherwise arise from mismodeling the \ac{CBC} population.
    }
    \label{fig:O5_GP}
    \script{pm_H0_twopanel.py}
\end{figure*}

In practice, the source-frame masses of individual \ac{GW} signals are not known~\citep[unless tidal information is available, e.g.][]{messenger_measuring_2012, 2021PhRvD.104h3528C}.
It is, however, possible to consider the \textit{population} of compact binaries at large and use known features in their source-frame mass distribution to obtain self-calibrated redshift estimates.
The full mass distribution therefore acts analogously to an electromagnetic spectrum, in which the apparent locations of spectral features relative to their rest-frame locations provide a redshift measurement.
The method of using the mass distribution of \ac{GW} sources to measure cosmological parameters has therefore been coined ``spectral sirens'' \citep{ezquiaga_spectral_2022}.
Spectral sirens were first demonstrated to be a feasible method to measure the Hubble constant by \cite{chernoff_gravitational_1993} and \cite{Taylor:2011fs} using the binary neutron star mass distribution, and extended to the \ac{BBH} mass distribution by \cite{farr_future_2019}. 
Spectral siren analyses have since been implemented by the LIGO--Virgo--KAGRA Collaborations using the latest \ac{GW} catalog \citep{gwtc3_cosmo}.

Central to the spectral siren methodology is knowledge of the compact binary mass distribution.
However, first principles models for mass distributions of merging compact binaries are not available: significant theoretical uncertainties exist about the shapes, locations, and very existence of predicted features in the mass distribution~\citep[e.g.][]{ zevin_constraining_2017, mapelli_binary_2020,2021ApJ...910..152Z,marchant_evolution_2023}.
This includes uncertainties about potentially dominant features, such as the existence of a ``pile-up'' due to pulsational-pair instability, a pair-instability-driven upper mass gap~\citep{farmer_mind_2019}, the maximum neutron star mass~\citep{fryer_theoretical_2001, alsing_evidence_2018}, and the existence of a putative lower mass gap between neutron stars and black holes~\citep{ozel_black_2010, farr_mass_2011, farah_bridging_2022}.
To this end, spectral siren cosmology relies on \textit{simultaneously} measuring a source-frame mass spectrum alongside cosmological parameters.
This is typically accomplished by adopting a phenomenological, parametric model for the mass distribution, usually composed of power laws and Gaussians \citep[e.g.][]{gwtc3_cosmo,mastrogiovanni_icarogw_2023}.

Such parametric modeling of the compact binary mass distribution raises its own set of dangers.
It is well known that different parametric models can generically yield very different constraints on cosmological parameters ~\citep{abbott_population_2021,abbott_population_2023}.
This is problematic: because the measured mass distribution serves as the template by which to extract redshifts, a mismodeled mass distribution would introduce systematic errors in inferred redshifts and, in turn, systematically bias any resulting cosmological inference~\citep{ezquiaga_spectral_2022,Mukherjee:2021rtw,mastrogiovanni_importance_2021,pierra_study_2023}.

The situation is demonstrated in Figure~\ref{fig:O5_GP}, in which we perform spectral siren cosmology on a simulated population of binary black holes.
We reconstruct the mass distribution using two parametric models, one that contains the true simulated mass distribution and one that does not.
While the former yields a measurement of $H_0$ consistent with the true underlying value, the latter does not.
Such systematic biases may already be relevant, as cosmological measurements by the LIGO-Virgo-KAGRA are known to depend on the choice of mass model used~\citep{gwtc3_cosmo}.
Furthermore, these biases may become a dominant source of uncertainty in the near future~\citep{pierra_study_2023}.
The prospects of such a dominant systematic uncertainty is troubling.
If prior knowledge of the mass distribution's morphology (whether an exact theoretical prediction or knowledge of the correct parametric family of models) is a prerequisite for the spectral siren method, the effectiveness of such a technique would be significantly hampered.

In this work, we explicitly demonstrate that no prior knowledge of the shape of the \ac{CBC} mass spectrum is necessary to use the spectral siren methodology. 
We do this by inferring $H(z)$ with a flexible, non-parametric model for the mass distribution of \acp{CBC} (blue shaded band in Figure~\ref{fig:O5_GP}). 
This model makes minimal prior assumptions about the shape of the mass distribution, enabling it to accurately infer a wide range of morphologies and remain agnostic to the astrophysical processes that give rise to features in the mass distribution.
Despite its flexibility, our approach is able to consistently obtain unbiased measurements of cosmological parameters, showing that non-parametric methods are not only sufficient for a spectral siren measurement, they can also mitigate systematic effects in the measurement caused by model misspecification.

The non-parametric mass model's ability to recover injected cosmological parameters demonstrates that the information in the spectral siren measurement does not come from the enforcement of specific features in the mass distribution.
Rather, it is provided by the assumption that either all \acp{CBC} follow a common mass distribution, or that any evolution of the mass distribution with redshift does not exactly mimic cosmology \citep{ezquiaga_spectral_2022}.

We find that our non-parametric model allows for a $%
  11.2
\unskip\label{output/nonparh0percent.txt}\unskip%
\%$ measurement of \Ho{} and a $%
  6.4
\unskip\label{output/Hz_percent.txt}\unskip%
\%$ measurement of $H(z=%
  0.9
\unskip\label{output/mostsensitivez.txt}\unskip%
)$ during \ac{O5}, when the detectors will reach their design sensitivity. 
We highlight measurements of \Ho{} within a $\Lambda$CDM universe in order to benchmark the accuracy and precision of our non-parametric method, as well as explore the role of spectral sirens in elucidating the Hubble tension.
However, a primary utility of spectral siren measurements will be in constraining $H(z)$ under different cosmological models, and at redshifts that are relatively inaccessible by electromagnetic observations, especially with next-generation gravitational-wave detectors \citep{You:2020wju, Ezquiaga:2020tns, ezquiaga_spectral_2022, Chen:2024gdn}.\footnote{See example spectral siren cosmological inference using parametric mass spectrum models: \href{https://github.com/ezquiaga/spectral_sirens}{https://github.com/ezquiaga/spectral\_sirens}.}

This paper is organized as follows: Section~\ref{sec:data generation} describes the simulated dataset.
Section~\ref{sec:ss} introduces the spectral siren method, demonstrating how cosmological parameters are inferred from the mass distribution of \ac{GW} sources.
Section~\ref{sec:model} describes the non-parametric mass distribution we develop for use within the spectral siren method.
In Section~\ref{sec:results}, we present the results of using parametric and non-parametric mass distributions, as well as projections for future constraints on $H(z)$.
We discuss the implications of our results and outline future work in Section~\ref{sec:discussion}.

This study was carried out using the reproducibility software \href{https://github.com/showyourwork/showyourwork}{\showyourwork} \citep{Luger2021}, which leverages continuous integration to programmatically download the data from \href{https://zenodo.org/}{zenodo.org}, create the figures, and compile the manuscript.
The icons next to each Figure caption are hyperlinks that lead to the code used to make that Figure (github icons), and to the data behind the Figure (cylindrical icons).
Some programatically-generated numbers in the text also have these icons, which function in the same way.
We encourage readers to click these hyperlinks to verify or reproduce the results and methods of this work.
The git repository associated to this study allows anyone to re-build the entire manuscript, and is publicly available at \GitHubURL{} and \url{https://zenodo.org/records/11199928}.

\section{Cosmology with an astrophysically-agnostic mass model}
\label{sec:methods}

The spectral siren method functions by identifying the relationship between luminosity distance and redshift that causes all source frame masses to follow a distribution that smoothly varies as a function of time. 
This allows for the simultaneous inference of both a mass distribution and cosmological parameters, even if the form of the source-frame mass distribution is not known in advance.
As described above, though, strongly parametrized models for the mass distribution yield biased measurements of cosmological parameters if they poorly approximate the true mass distribution of compact binaries.
We aim to circumvent such biases and instead model the population of \acp{GW} sources in a flexible and astrophysically-agnostic way.

There exist several non-parametric methods developed for this purpose \citep{tiwari_vamana_2021,edelman_aint_2022,sadiq_flexible_2022,rinaldi_hdpgmm_2022,edelman_cover_2023,mandel_extracting_2019,ray_non-parametric_2023,callister_parameter-free_2023}.
While well-suited to infer the \ac{GW} source population with a fixed cosmology, several of these methods employ fixed features in source frame mass, such as bin edges \citep{mandel_extracting_2019,ray_non-parametric_2023} or spline nodes \citep{edelman_aint_2022}.
Since these locations were chosen with a fixed cosmology, they risk causing the inference to prefer the cosmological parameters assumed when choosing the feature locations. 
Indeed, \citet{mastrogiovanni_importance_2021} show how using fixed features can significantly bias cosmological inference within the spectral siren methodology.
We therefore opt for a model of the source frame mass distribution that foregoes the need to define such features.

For this purpose, we construct a model with a \acf{GP}, a common tool for non-parametric inference. 
\Acp{GP} define a random space of functions in which any subset of function values are jointly Gaussian-distributed \citep{rasmussen_gaussian_2006}.
Their smoothness properties make them widely useful in \ac{GW} data analysis for regression problems, such as modeling time-domain waveforms \citep{doctor_statistical_2017, huerta_eccentric_2018} and the neutron star equation of state \citep{landry_nonparametric_2019}, density estimation problems, such as estimating posterior densities of single-event parameters from parameter estimation samples ~\citep{demilio_density_2021}, and as a prior on histogram bin heights for population inference \citep{mandel_model-independent_2017, li_flexible_2021, ray_non-parametric_2023}.

Our use case is slightly different from previous analyses:
we utilize a \ac{GP} as a prior on the functions that describe the primary mass distribution of \acp{CBC}.
This choice encodes very little prior information about the shape of the mass distribution, besides enforcing that it must be smooth.

\subsection{Simulated Data}
\label{sec:data generation}
To demonstrate the effectiveness of our \ac{GP}-based mass distribution in agnostically inferring both cosmological parameters and the population properties of \ac{GW} sources, we apply our methodology to a simulated dataset.
By generating a catalog of \ac{GW} sources from a known population and cosmological model, we are able to quantify the accuracy of our method.
The use of simulated data also enables us to make projections for future datasets and safely ignore dimensions such as spin that do not impact cosmological measurements but would otherwise be important to simultaneously fit to avoid biases in population inference of real data, as real data exhibit correlations between distance, mass, and spin measurements~\citep{biscoveanu_sources_2021}.

We design our simulated catalog to match the characteristics of the data expected from one year of observation in \ac{O5}. 
The \acp{BBH} in this catalog are drawn from an underlying population described by the \plp{} mass distribution presented in \citet{talbot_measuring_2018} and used in \citet{abbott_binary_2019, abbott_population_2023}, and follow the redshift distribution presented in \citet{callister_shouts_2020}, with hyperparameters consistent with those found in \citet{abbott_population_2023}.
We assume the mass distribution does not evolve across the redshift range to which the \ac{O5} detectors will be sensitive. 
This assumption is consistent with current data \citep{fishbach_when_2021,van_son_redshift_2022,abbott_population_2023}, in which no redshift evolution of the black hole mass function is detected.
At the same time, a redshift-dependent mass function is a generic astrophysical prediction, due either to changing evolutionary environments or evolving mixture fractions between distinct compact binary formation channels~\citep{neijssel_effect_2019,van_son_redshift_2022,2024arXiv240114837T,2024arXiv240212444Y}.
We discuss this possibility further in Section~\ref{sec:discussion}, but leave it primarily for future work.

We use the \texttt{GWMockCat}~ \citep{farah_things_2023} package to apply \ac{O5}-like selection effects to the drawn \acp{BBH}, generate realistic measurement uncertainty, and produce sensitivity estimates that are consistent with the simulated \ac{GW} signals.
We will use the term ``event'' to refer to \ac{GW} signals that pass the criteria for detection.
This process results in a catalog of $N_{\text{ev}} = %
  591
\unskip\label{output/num_found_events.txt}\unskip%
$ \ac{GW} signals that pass the criteria for detection, hereon called events.
Additional details of the data simulation, including the form of the injected population, are described in Appendix~\ref{ap:data generation}. 

\subsection{The Spectral Siren Method}
\label{sec:ss}
To simultaneously infer cosmological parameters and the population of \ac{GW} sources, we employ a hierarchical Bayesian analysis.
This allows us to undo the selection effects of \ac{GW} detectors to obtain a true, astrophysical population and constrain the cosmic expansion history.

Given a source population and background cosmology described by hyperparameters $\Lambda$, the likelihood of observing data $\{d\}$ that contains $N_{\text{ev}}$ detected \ac{GW} signals, each with parameters $\theta_i$, is \citep{loredo_handling_2009, Taylor:2011fs, mandel_extracting_2019,vitale_inferring_2020}
\begin{equation}
\begin{aligned}
    &p(\{d\},\{\theta\}|\Lambda) \propto \\
    &\hspace{1.2cm} e^{-N_{\text{exp}}(\Lambda)}\prod_i^{N_{\text{ev}}} p(d_i|\theta_i) \frac{\diff N}{\diff t_{\det} \diff \theta} (\theta_i;\Lambda) \, .
\end{aligned}
\label{eq:inhomog-poisson}
\end{equation}
Here, $\Lambda$ is the set of population parameters hyper-parameters and
$\frac{\diff N}{\diff t_{\det} \diff \theta} (\theta;\Lambda)$ is the detector frame merger rate density of \acp{BBH}, conditioned on hyper-parameters $\Lambda$. 
Following \citet{callister_parameter-free_2023}, we use a semicolon to explicitly indicate that this is a function of $\Lambda$, not a density over $\Lambda$.
$N_{\text{exp}}(\Lambda)$ is the expected number of detections given $\Lambda$ and the \ac{GW} detector sensitivity, and is calculated using a Monte Carlo sum over $N_{\text{inj}}$ found signals injected into the data stream \citep[see][for a detailed explanation of this process]{essick_estimating_2021, essick_precision_2022}.

In this work, we have restricted our analysis to the \ac{BBH} primary mass distribution.
However, the method can be trivially extended to the full mass distribution of \acp{CBC} \citep[e.g.][]{fishbach_does_2020, ezquiaga_spectral_2022}.
\new{Additionally, Appendix~\ref{ap:mass ratio} demonstrates that including the mass ratio distribution in the fit does not change the results of the analysis presented here.}

Since parameters of individual events are not perfectly measured, we marginalize over the possible properties of each event.
Practically, this is done by a Monte Carlo average over the posterior samples $\{\theta_j\}_i$ of each event $i$ and dividing out the prior used when inferring those posterior samples, $\pi_{\rm PE}(\theta)$:
\begin{equation}
\begin{aligned}
    &p(\{d\}|\Lambda) \\
    &\hspace{3mm} \propto e^{-N_{\text{exp}}(\Lambda)}\prod_i^{N_{\text{ev}}} \int 
    d\theta_i\, p(d_i|\theta_i) \frac{\diff N}{\diff t_{\det} \diff \theta} (\theta_i;\Lambda) \\
     &\hspace{3mm} \approx e^{-N_{\text{exp}}(\Lambda)}\prod_i^{N_{\text{ev}}} \frac{1}{N_{\rm samps}} \sum_{j=1}^{N_{\rm samps}} \frac{\frac{\diff N}{\diff t_{\det} \diff \theta} (\theta_{j,i};\Lambda)}{\pi_{\rm PE}(\theta_{j,i})}\,.
\end{aligned}
\label{eq:single-event-likelihood}
\end{equation}
When combined with a prior $p(\Lambda)$ (to be discussed below) on the population and cosmological parameters, the result is a posterior on both the compact binary population and the background cosmology. The full set of hyper-parameters $\Lambda$ therefore includes the shape of the mass distribution as well as all cosmological parameters that dictate the $D_L$--$z$ relation: the local expansion rate \Ho, the present fractional energy densities of dark matter \Omm, dark energy $\Omega_\Lambda$, and radiation $\Omega_r$, and the equation of state of dark energy $w$.
In this work, we fix $\Omega_\Lambda=1-\Omega_M, \Omega_r=0$ and $w=-1$ and use uniform priors on \Ho{} and \Omm{}, corresponding to a flat $\Lambda$CDM cosmology.

The process by which we sample the likelihood in Equation~\ref{eq:inhomog-poisson} is outlined in Appendix~\ref{ap:GP}.

\subsection{\Acl{GP}-based mass distribution}
\label{sec:model}

In this Section, we give an overview of the non-parametric mass model developed for this work.
Further details on this model, including an introduction to \acp{GP} and a discussion of their properties is given in Appendix~\ref{ap:GP}. 

\begin{figure*}
    \centering
    \includegraphics[width=\textwidth]{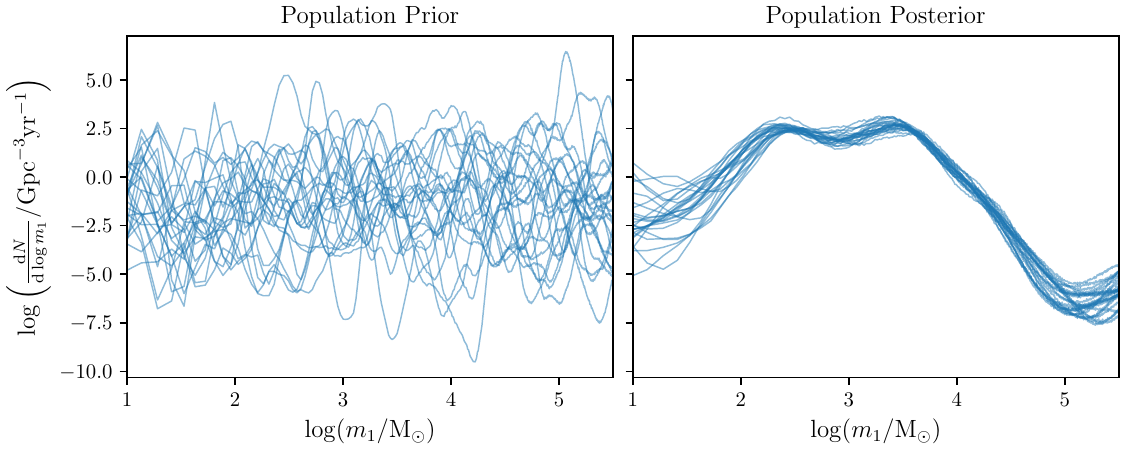}
    \caption{Draws from the \acf{GP} used to model the mass distribution.
    The left panel shows prior draws from the \ac{GP} and the right panel shows posterior draws once the population inference is performed on the simulated data.
    The posterior draws in the right panel are a subset of those used to create the 90\% credible intervals in Figure~\ref{fig:O5_GP}.}
    \label{fig:GP example}
    \script{GP_example_plot.py}
\end{figure*}

With the \ac{GP} approach, the hyper-parameters describing the mass distribution are the rate at each event-level posterior sample's source frame mass, and the rate at each found injection's source frame mass.
The \ac{GP} \emph{is} $p(\Lambda)$, the prior on population parameters (except in the case of cosmological parameters, which all have uniform priors).
This is demonstrated in Figure~\ref{fig:GP example}, where the left panel shows draws from a \ac{GP}, which are prior draws for the population inference.
The smooth appearance of individual draws from the population prior, as well as the absence of overdensities at specific source frame mass values in the full prior distribution shown in Figure~\ref{fig:GP example} demonstrate that we have successfully fulfilled our goal to construct a model without predefined features in source-frame mass.
Combined with the population likelihood in Equation~\ref{eq:inhomog-poisson} the prior illustrated in the left panel of Figure~\ref{fig:GP example} gives the population posterior in the right panel.

The lack of data just below the minimum black hole mass and just above the maximum black hole mass, combined with the fact that \ac{GW} detectors are sensitive to objects at those masses, causes the \ac{GP} to learn a relatively low merger rate at the edges of the mass distribution. 
On the other hand, there is both a lack of data and little detector sensitivity at masses above $\sim100\Msun$ and below $\sim5\Msun$.
The mass distribution is therefore uninformed in this region and the \ac{GP} reverts to its prior distribution, which resembles random scatter around the mean differential merger rate.
Similar effects can be seen in other non-parametric methods \citep{edelman_cover_2023, callister_parameter-free_2023}.
The combination of these two effects results in what appears as an uptick in the merger rate below $\sim5\Msun$ and above $\sim100\Msun$.
However, this reversion to the prior is uninformative for the \Ho{} constraint and does not affect inference on cosmological parameters.
Additionally, the posterior on \Ho{} is distinct from its prior distribution (uniform in the range $[30\Hunits,120\Hunits]$, indicating that the data is informative despite the flexibility of the population model.

We note that other non-parametric methods may be adapted to avoid predefined features, such as fitting for the locations of their features simultaneously with the rest of the inference \citep[e.g.][]{tiwari_vamana_2021} or, in the case of splines, by using a smoothing function that allows for features to occur at arbitrary locations appropriate smoothing \citep[e.g.][]{edelman_cover_2023}. 

The smoothness of a given \ac{GP} is determined by its kernel, which is a function that defines the covariance between input points in the \ac{GP} (in our case, two source frame mass values). 
It defines the notion of similarity between adjacent points and thereby encodes our assumptions about the smoothness of the source frame mass distribution \citep{rasmussen_gaussian_2006}.
Kernels themselves have parameters that determine their properties.
In our use case, these are one level further removed from hyper-parameters, so we adopt the terminology used in \citet{callister_parameter-free_2023} and call them ``hyper-hyper-parameters.''
We fit these hyper-hyper-parameters along with the hyper-parameters $\Lambda$ to minimize prior assumptions about the form of the mass distribution.

\section{Results}
\label{sec:results}
In this Section we show that fitting an incorrect functional form to the mass distribution of \acp{CBC} biases the inference of cosmological parameters when using the spectral siren methodology.
We then demonstrate that our flexible model alleviates this bias without the need to know the morphology of the mass distribution \emph{a priori}.
We illustrate this explicitly by using three different models for the source-frame mass distribution to infer the cosmic expansion rate from the simulated catalog described in Section~\ref{sec:data generation} and Appendix~\ref{ap:data generation}.
The three models are as follows:
\begin{enumerate}
    \item \plp{}, which includes the true mass distribution within its hyper-prior,
    \item the \bpl{} model presented in \citet{abbott_population_2021}, as we do not employ a high-mass truncation, which does not include the true mass distribution within its hyper-prior, and 
    \item the flexible, \ac{GP}-based model described in Section~\ref{sec:model}, which is able to closely approximate the morphology of the true mass distribution, along with many other morphologies.
\end{enumerate}
For all models considered in this work, we assume the form of the redshift distribution used to generate the data, described in Equation~\ref{eq:underlying redshift dist}.
We have performed the analysis both with a fixed redshift distribution and while simultaneously fitting for the redshift distribution and find no qualitative differences in our conclusions: fitting for the redshift distribution broadens the inferred posteriors on \Ho{} equally for all mass models, but does not affect their mean values.

The results of fitting each model to the same dataset are shown in Figure~\ref{fig:O5_GP}.
The left panel shows the inferred source-frame mass distribution for each of the considered models, and the right panel shows the corresponding posteriors on \Ho{}. 
We indicate the true underlying source mass distribution and \Ho{} value with solid black lines in each panel.

The fits presented in Figure~\ref{fig:O5_GP} are representative results from a single simulated catalog.
These provide insight into the full statistical results presented below.
In particular, it can be seen that the \bpl{} (orange curve) is inconsistent with the true value of $H_0$: in the run shown in Figure~\ref{fig:O5_GP}, the true value of \Ho{} is offset from the mean of the posterior by $%
  2.7
\unskip\label{output/BPLh0offset.txt}\unskip%
\sigma$.
In contrast, the \plp{} and \ac{GP}-based models (green and blue curves) are consistent with the underlying truth.
These models recover mean values of \Ho{} that are offset from the true value at $%
  0.8
\unskip\label{output/PLPh0offset.txt}\unskip%
\sigma$ and $%
  0.3
\unskip\label{output/nonparh0offset.txt}\unskip%
\sigma$, respectively. 
Additionally, the mass distribution inferred with the \ac{GP}-based model closely resembles the true, simulated distribution.
This indicates that using models that cannot accurately approximate the true mass distribution will lead to a noticeable systematic bias in the estimation of cosmological parameters.

This bias is not due to the need to know the morphology of the mass distribution \emph{a priori}, as the \ac{GP}-based model recovers the correct value of \Ho{} despite making minimal assumptions about the mass distribution.
In reality, we do not know the true functional form of the mass distribution, 
so it may be desirable to use a non-parametric approach to avoid potential systematic errors introduced by choosing a parametric model that likely does not contain the true mass distribution within its hyper-prior.

To obtain a quantitative measure of the systematic bias introduced by mis-modeling the mass distribution, we repeat the parametric analyses with 50 separate simulated catalogs of $\sim1,000$ events each.
We find that the \bpl{} model produces an over- or under-estimate of \Ho{} at greater than $1\sigma$ $%
  90
\unskip\label{output/BPL_bias_percent.txt}\unskip%
\%$ of the time, 
and the \plp{} model reaches the same level of bias only $%
  26
\unskip\label{output/PLP_bias_percent.txt}\unskip%
\%$ 
of the time, \new{meaning that the \bpl{} model produces a bias more than three times as often as the \plp{} model.}
\new{Additionally, we show in Appendix~\ref{ap:parametric bias} that the \bpl{} model typically overestimates \Ho{} whereas the \plp{} model produces a roughly equal number of over- and underestimates of \Ho{}.
This demonstrates that mis-modeling the mass distribution can introduce statistically significant systematic biases into measurements of cosmological parameters.}

Collectively, our results indicate that \emph{prior knowledge of the shape of the mass distribution is not required to perform an unbiased spectral siren measurement, so long as strong assumptions about the shape of the mass distribution are not made}.

\subsection{Projections for Future Measurements}
\begin{figure*}
    \centering
    \includegraphics[width=\textwidth]{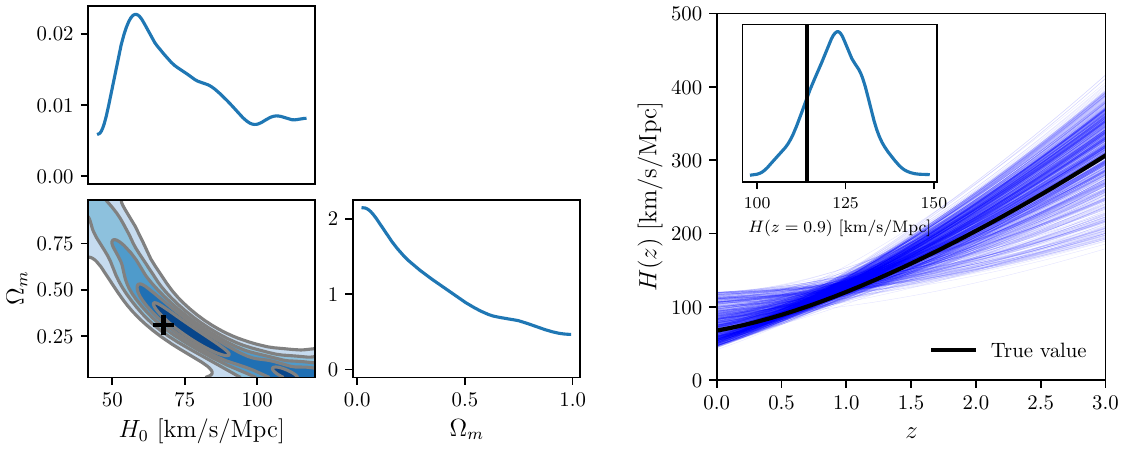}
    \caption{
    Projected constraints on multiple cosmological parameters after one year of observing at the \ac{LVK}'s design sensitivity, using the \acl{GP}-based spectral siren method.
    The right panel shows the inferred expansion history of the universe, $H(z)$.
    It will be measured most precisely at $z=%
  \input{output/mostsensitivez.txt}\unskip\label{output/mostsensitivez.txt}\unskip%
$, as can be seen by the narrowing of the inferred $H(z)$ curves there.
    The inset shows the posterior on $H(z=%
  \input{output/mostsensitivez.txt}\unskip\label{output/mostsensitivez.txt}\unskip%
)$.
    Black solid lines indicate the true value of $H(z)$ in both the inset and main panel.
    The left panel shows the two-dimensional posterior on \Ho{} and \Omm, with the true value indicated by a black ``+''.
    The two parameters are strongly degenerate because of the multiple ways of measuring $z=%
  \input{output/mostsensitivez.txt}\unskip\label{output/mostsensitivez.txt}\unskip%
$.
    Spectral sirens are particularly well suited for measuring cosmological parameters that affect the Universe at $z\gtrsim0.2$. 
    }
    \label{fig:O5_corner}
    \script{nonparametric_corner.py}
\end{figure*}

Figure~\ref{fig:O5_GP} demonstrates an expected $%
  11.2
\unskip\label{output/nonparh0percent.txt}\unskip%
\%$ ($1\sigma$ uncertainties) measurement of \Ho{} after one year of \ac{O5} using the \ac{GP}-based spectral siren method, and a $%
  8
\unskip\label{output/PLPh0percent.txt}\unskip%
\%$ measurement with parametric spectral sirens, demonstrating comparable statistical uncertainties.
However, we note that the precision reached in \ac{O5} may lessen depending on the actual level of measurement uncertainty in individual events, and the existence (or lack thereof) of a maximum mass feature, which we have assumed to be present in our simulated dataset.
These numbers are estimated from a fit to a single simulated catalog, but we find similar levels of statistical uncertainty from fits to different catalog realizations.
By the time of \ac{O5}, the \ac{GW} detector network is projected to detect \acp{BBH} up to redshift $\approx 3$, with most sources lying near redshift $\simeq 1.2$ \citep{chen_distance_2021}.
Additionally, next-generation detectors will be sensitive to sources up to redshift $\sim 100$.
This means that future \ac{GW} observations will be more sensitive to $H(z\gtrsim1)$ than to \Ho, and can therefore constrain several additional cosmological parameters \citep{Chen:2024gdn}.
We demonstrate this by repeating the same \ac{GP}-based spectral siren analysis while also simultaneously fitting for the local matter density, \Omm.
The result is shown in Figure~\ref{fig:O5_corner}.
We emphasize that these precise measurements over a wide range in redshift enable precision estimation of additional cosmological parameters governing $H(z)$.

We find \ac{O5} observations to be most sensitive to $H(z=%
  0.9
\unskip\label{output/mostsensitivez.txt}\unskip%
)$, which is measured at $%
  6.4
\unskip\label{output/Hz_percent.txt}\unskip%
\%$.
The left panel of Figure~\ref{fig:O5_corner} demonstrates a strong anti-correlation between the \Omm{} and \Ho{} posteriors, resulting in similarly informative constraints on the two parameters.
This is because \Ho{} controls the $y$-intercept of the $H(z)$ curves on the right panel, while \Omm{} informs the slope of those curves; the same measurement of $H(z\neq0)$ can be obtained by increasing the slope while decreasing the $y$-intercept, and vice versa.
Similar behavior can be observed in current measurements of the \ac{BBH} redshift distribution, which exhibits a tightening of the posterior at $z\sim0.2$ with current observations \citep{abbott_population_2023, callister_parameter-free_2023}.

Next-generation detectors will be sensitive to a larger range of redshifts \citep{et_steering_committee_einstein_2020, evans_horizon_2021}, and will therefore break the degeneracy between cosmological parameters and allow for tighter constraints on both \Omm{} and \Ho{}.
However, the small cosmological volume (and thus low number of mergers) at low redshift will generally limit the constraining power of spectral sirens at $z=0$, potentially making this method more sensitive to cosmological parameters that affect higher redshifts. 
Combining spectral sirens with other methods that are sensitive to the local expansion rate, such as those that employ electromagnetic counterparts, may increase the precision of \ac{GW} standard sirens at all redshifts \citep[e.g.][]{Chen:2024gdn}.

%

\section{Discussion}
\label{sec:discussion}

\acp{GW} are unique cosmic messengers in that they carry both redshift and distance information, making them remarkably clean probes of the Universe's expansion history calibrated directly by the theory of general relativity. 
However, the current method of determining \ac{GW} redshifts via the distribution of their source masses (i.e. spectral sirens) employs an assumption on the shape of their population, typically encapsulated by simplified parametric functions. 
This choice, often presumed to be necessary or fundamental to the method, may introduce a systematic bias to a measurement of the cosmological parameters that is otherwise appealing for its unique elegance and simplicity.
    
In this work we show that a specific choice of mass distribution is unnecessary to arrive at informative and accurate posteriors on \Ho{} and \Omm{}.
We do this by accurately measuring these cosmological parameters with a highly flexible model for the mass distribution. 
This reinforces the notion that the information in the spectral siren measurement comes from the assumption that all \ac{GW} sources come from the same population, a far less stringent statement than the assumption that we understand the exact astrophysical processes which give rise to that population (i.e. the physics governing compact binary formation and evolution).

Spectral sirens are particularly useful in the context of the current Hubble tension: a disagreement between multiple methods of measuring the local expansion rate of the universe \citep{freedman_measurements_2021}.
As this tension can only be explained by non-standard physics or yet-unknown systematic uncertainties in either the cosmic distance ladder or cosmic microwave background measurements, direct and independent probes of the local expansion---such as those presented here---may help determine whether the current discrepancy represents a fundamental crack in our standard model of physics and/or cosmology.
    
Neither the mechanism from which compact binaries were formed, nor the physical processes within the mechanism, have been conclusively determined.
The true functional form of the mass distribution therefore remains elusive, meaning that systematic uncertainties arising from an incorrect choice for the form of the mass distribution are inevitable.
With current observations, these systematic effects are likely smaller than statistical uncertainties.
However, next-generation detectors will herald sufficient \ac{GW} observations to substantially decrease statistical uncertainty in these measurements; for example, \citet{pierra_study_2023} show that incorrect assumptions about the shape of the mass distribution can lead to $\sim3\sigma$ systematic biases in \Ho{} with catalogs of 2,000 events, although this bias may be an overestimate as it does not include measurement uncertainty of the \ac{GW} parameters.
Thus, non-parametric approaches may be preferable to avoid the systematic errors associated with choosing a parametric model.

In parallel with this work, \citet{magana_hernandez_beyond_2024} performed a spectral siren analysis on public \ac{LVK} data using histogram bins defined at fixed locations in source-frame mass to flexibly measure the \ac{BBH} mass distribution \citep[as originally presented in][]{mandel_model-independent_2017,ray_non-parametric_2023}.
Both our method and theirs employ \acp{GP}.
In our approach, the black hole mass function is \textit{itself} described by a Gaussian process, with no predefined features.
In \citet{magana_hernandez_beyond_2024}, on the other hand, the mass distribution is fundamentally a binned piecewise constant function, with a \ac{GP} prior governing the relative heights of these bins.
While this binned model is flexible, sharp bin edges at predefined locations nevertheless constitute unphysical features.
These features may yield overly optimistic constraints on cosmological parameters, possibly explaining the increase in precision that \citet{magana_hernandez_beyond_2024} find relative to the parametric analysis on the exact same dataset performed in \citet{gwtc3_cosmo}.
\citet{magana_hernandez_beyond_2024} also include a binned reconstruction of the \ac{BBH} merger rate as a function of redshift, which we fix or model parametrically in our own work.
We stress, though, that this is not an intrinsic limitation of our method; one could straightforwardly adopt a second Gaussian process to govern the overall scaling of the merger rate with redshift.

More interesting is the possibility that the black hole mass distribution itself evolves with redshift.
As demonstrated in \citet{ezquiaga_spectral_2022}, the degeneracy between an evolving mass distribution and the expansion of the universe can be broken provided that there are multiple features present in the mass distribution, and that we do not live in a fine-tuned universe where the evolutionary effects governing the \ac{CBC} mass distribution perfectly mimick the effects of cosmological redshift.
The first condition (multiple features) is known to be met in current data \citep{abbott_population_2023}, with three robust features \citep{farah_things_2023}: a maximum black hole mass and overdensities at $\sim10\Msun$ and $\sim35\Msun$.
It is also predicted by many population synthesis studies, \citep[e.g.][]{mapelli_cosmic_2018,giacobbo_merging_2018,van_son_locations_2023}.
The latter scenario---with features identically and monotonically shifting to higher masses with increasing redshift---would be astrophysically unlikely as the locations of features in the mass distribution are each thought to be governed by fundamentally different physical processes \citep{mapelli_binary_2020}.
Extensions to the non-parametric methods presented here will allow for arbitrary redshift evolution.
Additionally, since \aclp{GP} naturally scale to multiple data dimensions \citep{rasmussen_gaussian_2006}, the method presented here can easily be generalized to fit the redshift dependence of the mass distribution. 
Future work will extend the method developed here to mass distributions that are allowed to evolve with redshift.
When applying an evolving mass distribution model to a dataset that exhibits evolution, we expect errors on cosmological parameters to broaden.
Noticeable redshift evolution in the mass distribution is expected to occur above $z\simeq 1.5$, \citep{van_son_redshift_2022}, which is not yet relevant in current data or in expected O5 data, but will certainly be visible in next-generation detectors.

Because \ac{GW} observations are the only data input to spectral sirens, they are sensitive to the expansion history of the universe over a wide range of cosmological redshift, rather than just the local expansion \Ho.
We have shown this by simultaneously measuring \Omm{} and \Ho{}; the method can be trivially expanded to constrain additional cosmological parameters that govern $H(z)$, such as the dark energy equation of state parameter, $w$. 
It is also possible to use a non-$\Lambda$CDM cosmological model for $H(z)$, with different parameters entirely.
It may also be possible to forego the need for a parametric representation of $H(z)$ altogether, for example by using bins in luminosity distance.
A similar technique using redshift bins is commonly used in cosmological measurements using galaxy surveys and type 1a supernovae \citep[e.g.][]{perlmutter_measuring_2003,anderson_clustering_2014,aghamousa_desi_2016}, but we have not pursued that possibility in this work.
Such a method would likely increase uncertainties on $H(z)$, especially for $z\gtrsim 1.5$.
We find that for \acl{O5}, spectral sirens will be most constraining at $z=%
  0.9
\unskip\label{output/mostsensitivez.txt}\unskip%
$.
This redshift is larger than the expected redshifts of detectable electromagnetic counterparts of binary neutron star mergers~\citep{kiendrebeogo_updated_2023}, implying that upgrades to current \ac{GW} detectors will allow the spectral siren method to probe $H(z)$ at otherwise unexplored distances by \acp{GW}. 

Proposed next-generation \ac{GW} detectors such as Cosmic Explorer and Einstein Telescope will be sensitive to \acp{CBC} across cosmic time (out to $z\sim100$)~\citep{et_steering_committee_einstein_2020, evans_horizon_2021}.
Future cosmological surveys, such as those enabled by the Nancy Grace Roman Space Telescope and the Vera Rubin Observatory, are expected to be able to precisely measure $H(z)$ to $z\sim3$ \citep{spergel_wide-field_2015}, making the spectral siren method uniquely positioned to measure the expansion of the universe at high redshift.
When combined with bright and dark siren methods, the low-redshift expansion history will also be well-constrained \citep{Chen:2024gdn}.
Sensitivity to high redshifts is a feature of the spectral siren method in general, and non-parametric methods such as the one presented here will be imperative to avoid systematic biases in spectral siren cosmology. 

\begin{acknowledgments}
    The authors thank Reed Essick, Ben Farr, Maya Fishbach, Utkarsh Mali, and Colm Talbot for helpful conversations. 
    A.M.F. is supported by the National Science Foundation Graduate Research Fellowship Program under Grant No. DGE-1746045.
    J.M.E. is supported by the European Union’s Horizon 2020 research and innovation program under the Marie Sklodowska-Curie grant agreement No. 847523 INTERACTIONS, and by VILLUM FONDEN (grant no. 53101 and 37766). 
    M.Z. gratefully acknowledges funding from the Brinson Foundation in support of astrophysics research at the Adler Planetarium.
    D.E.H is supported by NSF grants AST-2006645 and PHY-2110507, as well as by the Kavli Institute for Cosmological Physics through an endowment from the Kavli Foundation and its founder Fred Kavli.
    The Tycho supercomputer hosted at the SCIENCE HPC center at the University of Copenhagen was used for supporting this work.
    This material is based upon work supported by NSF's LIGO Laboratory which is a major facility fully funded by the National Science Foundation.
\end{acknowledgments}

\software{\texttt{numpyro} \citep{phan_composable_2019,bingham_pyro_2019}, \texttt{tinygp} \citep{foreman-mackey_tinygp_2021}, \texttt{arviz} \citep{kumar_arviz_2019}, \texttt{jax}, \showyourwork \citep{Luger2021}
          }

\bibliography{bib}

\begin{thebibliography}{}
\expandafter\ifx\csname natexlab\endcsname\relax\def\natexlab#1{#1}\fi
\providecommand{\url}[1]{\href{#1}{#1}}
\providecommand{\dodoi}[1]{doi:~\href{http://doi.org/#1}{\nolinkurl{#1}}}
\providecommand{\doeprint}[1]{\href{http://ascl.net/#1}{\nolinkurl{http://ascl.net/#1}}}
\providecommand{\doarXiv}[1]{\href{https://arxiv.org/abs/#1}{\nolinkurl{https://arxiv.org/abs/#1}}}

\bibitem[{Aasi {et~al.}(2015)Aasi, Abbott, Abbott, Abbott, Abernathy,
  {et~al.}}]{aasi_advanced_2015}
Aasi, J., Abbott, B.~P., Abbott, R., {et~al.} 2015, Classical and Quantum
  Gravity, 32, 074001, \dodoi{10.1088/0264-9381/32/7/074001}

\bibitem[{Abbott {et~al.}(2019)Abbott, Abbott, Abbott, Abraham, Acernese,
  {et~al.}}]{abbott_binary_2019}
Abbott, B.~P., Abbott, R., Abbott, T.~D., {et~al.} 2019, The Astrophysical
  Journal Letters, 882, L24, \dodoi{10.3847/2041-8213/ab3800}

\bibitem[{Abbott {et~al.}(2020{\natexlab{a}})Abbott, Abbott, Abbott, Abraham,
  Acernese, {et~al.}}]{obsscen_noise_curves}
---. 2020{\natexlab{a}}, {LIGO}-{T2000012}-v2: {Noise} curves used for
  {Simulations} in the update of the {Observing} {Scenarios} {Paper}.
\newblock \url{https://dcc.ligo.org/LIGO-T2000012-v2/public}

\bibitem[{Abbott {et~al.}(2020{\natexlab{b}})Abbott, Abbott, Abbott, Abraham,
  Acernese, {et~al.}}]{abbott_prospects_2020}
---. 2020{\natexlab{b}}, Living Reviews in Relativity, 23, 3,
  \dodoi{10.1007/s41114-020-00026-9}

\bibitem[{Abbott {et~al.}(2017{\natexlab{a}})Abbott, Abbott, Abbott, Acernese,
  Ackley, {et~al.}}]{abbott_multi-messenger_2017}
---. 2017{\natexlab{a}}, The Astrophysical Journal, 848, L12,
  \dodoi{10.3847/2041-8213/aa91c9}

\bibitem[{Abbott {et~al.}(2017{\natexlab{b}})Abbott, Abbott, Abbott, Acernese,
  Ackley, {et~al.}}]{abbott_gravitational-wave_2017}
---. 2017{\natexlab{b}}, Nature, 551, 85, \dodoi{10.1038/nature24471}

\bibitem[{Abbott {et~al.}(2021{\natexlab{a}})Abbott, Abbott, Abbott, Abraham,
  Acernese, Ackley, Adams, Adhikari, {et~al.}}]{abbott_gravitational-wave_2021}
---. 2021{\natexlab{a}}, The Astrophysical Journal, 909, 218,
  \dodoi{10.3847/1538-4357/abdcb7}

\bibitem[{Abbott {et~al.}(2021{\natexlab{b}})Abbott, Abbott, Abraham, Acernese,
  Ackley, {et~al.}}]{abbott_population_2021}
Abbott, R., Abbott, T.~D., Abraham, S., {et~al.} 2021{\natexlab{b}}, The
  Astrophysical Journal Letters, 913, L7, \dodoi{10.3847/2041-8213/abe949}

\bibitem[{Abbott {et~al.}(2023)Abbott, Abbott, Acernese, Ackley, Adams,
  Adhikari, {et~al.}}]{abbott_population_2023}
Abbott, R., Abbott, T.~D., Acernese, F., {et~al.} 2023, Physical Review X, 13,
  011048, \dodoi{10.1103/PhysRevX.13.011048}

\bibitem[{{Abbott} {et~al.}(2023){Abbott}, {Abbott}, {Acernese}, {Ackley},
  {Adams}, {et~al.}}]{2023PhRvX..13d1039A}
{Abbott}, R., {Abbott}, T.~D., {Acernese}, F., {et~al.} 2023, Physical Review
  X, 13, 041039, \dodoi{10.1103/PhysRevX.13.041039}

\bibitem[{Abbott {et~al.}(2021{\natexlab{c}})Abbott, Abe, Acernese, Ackley,
  Adhikari, {et~al.}}]{gwtc3_cosmo}
Abbott, R., Abe, H., Acernese, F., {et~al.} 2021{\natexlab{c}}, Constraints on
  the cosmic expansion history from {GWTC}-3,  arXiv,
  \dodoi{10.48550/arXiv.2111.03604}

\bibitem[{Acernese {et~al.}(2014)Acernese, Agathos, Agatsuma, Aisa, Allemandou,
  {et~al.}}]{acernese_advanced_2014}
Acernese, F., Agathos, M., Agatsuma, K., {et~al.} 2014, Classical and Quantum
  Gravity, 32, 024001, \dodoi{10.1088/0264-9381/32/2/024001}

\bibitem[{Adachi \& Kasai(2012)}]{adachi_analytical_2012}
Adachi, M., \& Kasai, M. 2012, Progress of Theoretical Physics, 127, 145,
  \dodoi{10.1143/PTP.127.145}

\bibitem[{Aghamousa {et~al.}(2016)Aghamousa, Aguilar, Ahlen, Alam, Allen,
  Prieto, Annis, Bailey, Balland, Ballester, Baltay, Beaufore, Bebek, Beers,
  Bell, Bernal, Besuner, Beutler, Blake, Bleuler, Blomqvist, Blum, Bolton,
  Briceno, Brooks, Brownstein, Buckley-Geer, Burden, Burtin, Busca, Cahn, Cai,
  Cardiel-Sas, Carlberg, Carton, Casas, Castander, Cervantes-Cota, Claybaugh,
  Close, Coker, Cole, Comparat, Cooper, Cousinou, Crocce, Cuby, Cunningham,
  Davis, Dawson, de~la Macorra, De~Vicente, Delubac, Derwent, Dey, Dhungana,
  Ding, Doel, Duan, Ealet, Edelstein, Eftekharzadeh, Eisenstein, Elliott,
  Escoffier, Evatt, Fagrelius, Fan, Fanning, Farahi, Farihi, Favole, Feng,
  Fernandez, Findlay, Finkbeiner, Fitzpatrick, Flaugher, Flender, Font-Ribera,
  Forero-Romero, Fosalba, Frenk, Fumagalli, Gaensicke, Gallo, Garcia-Bellido,
  Gaztanaga, Fusillo, Gerard, Gershkovich, Giannantonio, Gillet, Gonzalez-de
  Rivera, Gonzalez-Perez, Gott, Graur, Gutierrez, Guy, Habib, Heetderks,
  Heetderks, Heitmann, Hellwing, Herrera, Ho, Holland, Honscheid, Huff,
  Hutchinson, Huterer, Hwang, Laguna, Ishikawa, Jacobs, Jeffrey, Jelinsky,
  Jennings, Jiang, Jimenez, Johnson, Joyce, Jullo, Juneau, Kama, Karcher,
  Karkar, Kehoe, Kennamer, Kent, Kilbinger, Kim, Kirkby, Kisner, Kitanidis,
  Kneib, Koposov, Kovacs, Koyama, Kremin, Kron, Kronig, Kueter-Young, Lacey,
  Lafever, Lahav, Lambert, Lampton, Landriau, Lang, Lauer, Goff, Guillou,
  Van~Suu, Lee, Lee, Leitner, Lesser, Levi, L'Huillier, Li, Liang, Lin, Linder,
  Loebman, Lukić, Ma, MacCrann, Magneville, Makarem, Manera, Manser, Marshall,
  Martini, Massey, Matheson, McCauley, McDonald, McGreer, Meisner, Metcalfe,
  Miller, Miquel, Moustakas, Myers, Naik, Newman, Nichol, Nicola, da~Costa,
  Nie, Niz, Norberg, Nord, Norman, Nugent, O'Brien, Oh, Olsen, Padilla,
  Padmanabhan, Padmanabhan, Palanque-Delabrouille, Palmese, Pappalardo, Pâris,
  Park, Patej, Peacock, Peiris, Peng, Percival, Perruchot, Pieri, Pogge,
  Pollack, Poppett, Prada, Prakash, Probst, Rabinowitz, Raichoor, Ree,
  Refregier, Regal, Reid, Reil, Rezaie, Rockosi, Roe, Ronayette, Roodman, Ross,
  Ross, Rossi, Rozo, Ruhlmann-Kleider, Rykoff, Sabiu, Samushia, Sanchez,
  Sanchez, Schlegel, Schneider, Schubnell, Secroun, Seljak, Seo, Serrano,
  Shafieloo, Shan, Sharples, Sholl, Shourt, Silber, Silva, Sirk, Slosar, Smith,
  Smoot, Som, Song, Sprayberry, Staten, Stefanik, Tarle, Tie, Tinker, Tojeiro,
  Valdes, Valenzuela, Valluri, Vargas-Magana, Verde, Walker, Wang, Wang,
  Weaver, Weaverdyck, Wechsler, Weinberg, White, Yang, Yeche, Zhang, Zhao,
  Zheng, Zhou, Zhou, Zhu, Zou, \& Zu}]{aghamousa_desi_2016}
Aghamousa, A., Aguilar, J., Ahlen, S., {et~al.} 2016, The {DESI} {Experiment}
  {Part} {I}: {Science},{Targeting}, and {Survey} {Design},  arXiv.
\newblock \url{http://arxiv.org/abs/1611.00036}

\bibitem[{Akutsu {et~al.}(2021)Akutsu, Ando, Arai, Arai, Araki,
  {et~al.}}]{akutsu_overview_2021}
Akutsu, T., Ando, M., Arai, K., {et~al.} 2021, Progress of Theoretical and
  Experimental Physics, 2021, 05A101, \dodoi{10.1093/ptep/ptaa125}

\bibitem[{Alsing {et~al.}(2018)Alsing, Silva, \& Berti}]{alsing_evidence_2018}
Alsing, J., Silva, H.~O., \& Berti, E. 2018, Monthly Notices of the Royal
  Astronomical Society, 478, 1377, \dodoi{10.1093/mnras/sty1065}

\bibitem[{Anderson {et~al.}(2014)Anderson, Aubourg, Bailey, Beutler, Bhardwaj,
  Blanton, Bolton, Brinkmann, Brownstein, Burden, Chuang, Cuesta, Dawson,
  Eisenstein, Escoffier, Gunn, Guo, Ho, Honscheid, Howlett, Kirkby, Lupton,
  Manera, Maraston, McBride, Mena, Montesano, Nichol, Nuza, Olmstead,
  Padmanabhan, Palanque-Delabrouille, Parejko, Percival, Petitjean, Prada,
  Price-Whelan, Reid, Roe, Ross, Ross, Sabiu, Saito, Samushia, Sanchez,
  Schlegel, Schneider, Scoccola, Seo, Skibba, Strauss, Swanson, Thomas, Tinker,
  Tojeiro, Magana, Verde, Wake, Weaver, Weinberg, White, Xu, Yeche, Zehavi, \&
  Zhao}]{anderson_clustering_2014}
Anderson, L., Aubourg, E., Bailey, S., {et~al.} 2014, Monthly Notices of the
  Royal Astronomical Society, 441, 24, \dodoi{10.1093/mnras/stu523}

\bibitem[{Bingham {et~al.}(2019)Bingham, Chen, Jankowiak, Obermeyer, Pradhan,
  Karaletsos, Singh, Szerlip, Horsfall, \& Goodman}]{bingham_pyro_2019}
Bingham, E., Chen, J.~P., Jankowiak, M., {et~al.} 2019, Journal of Machine
  Learning Research, 20, 1.
\newblock \url{http://jmlr.org/papers/v20/18-403.html}

\bibitem[{Biscoveanu {et~al.}(2021)Biscoveanu, Talbot, \&
  Vitale}]{biscoveanu_sources_2021}
Biscoveanu, S., Talbot, C., \& Vitale, S. 2021, arXiv:2111.13619 [astro-ph,
  physics:gr-qc].
\newblock \url{http://arxiv.org/abs/2111.13619}

\bibitem[{Callister {et~al.}(2020)Callister, Fishbach, Holz, \&
  Farr}]{callister_shouts_2020}
Callister, T., Fishbach, M., Holz, D.~E., \& Farr, W.~M. 2020, The
  Astrophysical Journal, 896, L32, \dodoi{10.3847/2041-8213/ab9743}

\bibitem[{Callister \& Farr(2023)}]{callister_parameter-free_2023}
Callister, T.~A., \& Farr, W.~M. 2023, A {Parameter}-{Free} {Tour} of the
  {Binary} {Black} {Hole} {Population},  arXiv,
  \dodoi{10.48550/arXiv.2302.07289}

\bibitem[{{Chatterjee} {et~al.}(2021){Chatterjee}, {Hegade K.~R.}, {Holder},
  {Holz}, {Perkins}, {Yagi}, \& {Yunes}}]{2021PhRvD.104h3528C}
{Chatterjee}, D., {Hegade K.~R.}, A., {Holder}, G., {et~al.} 2021, \prd, 104,
  083528, \dodoi{10.1103/PhysRevD.104.083528}

\bibitem[{Chen {et~al.}(2024)Chen, Ezquiaga, \& Gupta}]{Chen:2024gdn}
Chen, H.-Y., Ezquiaga, J.~M., \& Gupta, I. 2024.
\newblock \doarXiv{2402.03120}

\bibitem[{Chen {et~al.}(2018)Chen, Fishbach, \& Holz}]{chen_two_2018}
Chen, H.-Y., Fishbach, M., \& Holz, D.~E. 2018, Nature, 562, 545,
  \dodoi{10.1038/s41586-018-0606-0}

\bibitem[{Chen {et~al.}(2021)Chen, Holz, Miller, Evans, Vitale, \&
  Creighton}]{chen_distance_2021}
Chen, H.-Y., Holz, D.~E., Miller, J., {et~al.} 2021, Classical and Quantum
  Gravity, 38, 055010, \dodoi{10.1088/1361-6382/abd594}

\bibitem[{Chen {et~al.}(2019)Chen, Li, \& Cao}]{chen_mass-redshift_2019}
Chen, X., Li, S., \& Cao, Z. 2019, Monthly Notices of the Royal Astronomical
  Society, 485, L141, \dodoi{10.1093/mnrasl/slz046}

\bibitem[{Chernoff \& Finn(1993)}]{chernoff_gravitational_1993}
Chernoff, D.~F., \& Finn, L.~S. 1993, The Astrophysical Journal, 411, L5,
  \dodoi{10.1086/186898}

\bibitem[{Coulter {et~al.}(2017)Coulter, Foley, Kilpatrick, Drout, Piro,
  Shappee, Siebert, Simon, Ulloa, Kasen, Madore, Murguia-Berthier, Pan,
  Prochaska, Ramirez-Ruiz, Rest, \& Rojas-Bravo}]{coulter_swope_2017}
Coulter, D.~A., Foley, R.~J., Kilpatrick, C.~D., {et~al.} 2017, Science, 358,
  1556, \dodoi{10.1126/science.aap9811}

\bibitem[{Del~Pozzo(2012)}]{del_pozzo_inference_2012}
Del~Pozzo, W. 2012, Physical Review D, 86, 043011,
  \dodoi{10.1103/PhysRevD.86.043011}

\bibitem[{D'Emilio {et~al.}(2021)D'Emilio, Green, \&
  Raymond}]{demilio_density_2021}
D'Emilio, V., Green, R., \& Raymond, V. 2021, Monthly Notices of the Royal
  Astronomical Society, 508, 2090, \dodoi{10.1093/mnras/stab2623}

\bibitem[{Doctor {et~al.}(2017)Doctor, Farr, Holz, \&
  Pürrer}]{doctor_statistical_2017}
Doctor, Z., Farr, B., Holz, D.~E., \& Pürrer, M. 2017, Physical Review D, 96,
  123011, \dodoi{10.1103/PhysRevD.96.123011}

\bibitem[{Edelman {et~al.}(2022)Edelman, Doctor, Godfrey, \&
  Farr}]{edelman_aint_2022}
Edelman, B., Doctor, Z., Godfrey, J., \& Farr, B. 2022, The Astrophysical
  Journal, 924, 101, \dodoi{10.3847/1538-4357/ac3667}

\bibitem[{Edelman {et~al.}(2023)Edelman, Farr, \& Doctor}]{edelman_cover_2023}
Edelman, B., Farr, B., \& Doctor, Z. 2023, The Astrophysical Journal, 946, 16,
  \dodoi{10.3847/1538-4357/acb5ed}

\bibitem[{{Essick}(2023)}]{2023PhRvD.108d3011E}
{Essick}, R. 2023, \prd, 108, 043011, \dodoi{10.1103/PhysRevD.108.043011}

\bibitem[{Essick \& Farr(2022)}]{essick_precision_2022}
Essick, R., \& Farr, W. 2022, Precision {Requirements} for {Monte} {Carlo}
  {Sums} within {Hierarchical} {Bayesian} {Inference}, Tech. rep.
\newblock \url{https://ui.adsabs.harvard.edu/abs/2022arXiv220400461E}

\bibitem[{Essick \& Fishbach(2021)}]{essick_estimating_2021}
Essick, R., \& Fishbach, M. 2021, On {Estimating} {Rates} from {Monte}-{Carlo}
  {Integrals} over {Injection} {Sets}.
\newblock \url{https://dcc.ligo.org/T2000100}

\bibitem[{Essick \& Fishbach(2024)}]{essick_ensuring_2024}
---. 2024, The Astrophysical Journal, 962, 169,
  \dodoi{10.3847/1538-4357/ad1604}

\bibitem[{ET~steering committee(2020)}]{et_steering_committee_einstein_2020}
ET~steering committee, e.~a. 2020, Einstein {Telescope}: {Science} {Case},
  {Design} {Study} and {Feasibility} {Report}, Official document ET-0028A-20.
\newblock \url{https://apps.et-gw.eu/tds/?content=3&r=17196}

\bibitem[{Evans {et~al.}(2021)Evans, Adhikari, Afle, Ballmer, Biscoveanu,
  Borhanian, Brown, Chen, Eisenstein, Gruson, Gupta, Hall, Huxford, Kamai,
  Kashyap, Kissel, Kuns, Landry, Lenon, Lovelace, McCuller, Ng, Nitz, Read,
  Sathyaprakash, Shoemaker, Slagmolen, Smith, Srivastava, Sun, Vitale, \&
  Weiss}]{evans_horizon_2021}
Evans, M., Adhikari, R.~X., Afle, C., {et~al.} 2021, A {Horizon} {Study} for
  {Cosmic} {Explorer}: {Science}, {Observatories}, and {Community},
  \dodoi{10.48550/arXiv.2109.09882}

\bibitem[{Ezquiaga \& Holz(2021)}]{Ezquiaga:2020tns}
Ezquiaga, J.~M., \& Holz, D.~E. 2021, Astrophys. J. Lett., 909, L23,
  \dodoi{10.3847/2041-8213/abe638}

\bibitem[{Ezquiaga \& Holz(2022)}]{ezquiaga_spectral_2022}
---. 2022, Physical Review Letters, 129, 061102,
  \dodoi{10.1103/PhysRevLett.129.061102}

\bibitem[{Farah {et~al.}(2022)Farah, Fishbach, Essick, Holz, \&
  Galaudage}]{farah_bridging_2022}
Farah, A., Fishbach, M., Essick, R., Holz, D.~E., \& Galaudage, S. 2022, The
  Astrophysical Journal, 931, 108, \dodoi{10.3847/1538-4357/ac5f03}

\bibitem[{Farah {et~al.}(2023)Farah, Edelman, Zevin, Fishbach, María~Ezquiaga,
  Farr, \& Holz}]{farah_things_2023}
Farah, A.~M., Edelman, B., Zevin, M., {et~al.} 2023, The Astrophysical Journal,
  955, 107, \dodoi{10.3847/1538-4357/aced02}

\bibitem[{Farmer {et~al.}(2019)Farmer, Renzo, de~Mink, Marchant, \&
  Justham}]{farmer_mind_2019}
Farmer, R., Renzo, M., de~Mink, S.~E., Marchant, P., \& Justham, S. 2019, The
  Astrophysical Journal, 887, 53, \dodoi{10.3847/1538-4357/ab518b}

\bibitem[{Farr {et~al.}(2019)Farr, Fishbach, Ye, \& Holz}]{farr_future_2019}
Farr, W.~M., Fishbach, M., Ye, J., \& Holz, D.~E. 2019, The Astrophysical
  Journal, 883, L42, \dodoi{10.3847/2041-8213/ab4284}

\bibitem[{Farr {et~al.}(2011)Farr, Sravan, Cantrell, Kreidberg, Bailyn, Mandel,
  \& Kalogera}]{farr_mass_2011}
Farr, W.~M., Sravan, N., Cantrell, A., {et~al.} 2011, 741, 103,
  \dodoi{10.1088/0004-637X/741/2/103}

\bibitem[{Fishbach {et~al.}(2020)Fishbach, Essick, \&
  Holz}]{fishbach_does_2020}
Fishbach, M., Essick, R., \& Holz, D.~E. 2020, The Astrophysical Journal, 899,
  L8, \dodoi{10.3847/2041-8213/aba7b6}

\bibitem[{Fishbach {et~al.}(2019)Fishbach, Gray, Magaña~Hernandez, Qi, Sur,
  {et~al.}}]{fishbach_standard_2019}
Fishbach, M., Gray, R., Magaña~Hernandez, I., {et~al.} 2019, The Astrophysical
  Journal, 871, L13, \dodoi{10.3847/2041-8213/aaf96e}

\bibitem[{Fishbach \& Holz(2017)}]{fishbach_where_2017}
Fishbach, M., \& Holz, D.~E. 2017, The Astrophysical Journal, 851, L25,
  \dodoi{10.3847/2041-8213/aa9bf6}

\bibitem[{Fishbach {et~al.}(2021)Fishbach, Doctor, Callister, Edelman, Ye,
  Essick, Farr, Farr, \& Holz}]{fishbach_when_2021}
Fishbach, M., Doctor, Z., Callister, T., {et~al.} 2021, The Astrophysical
  Journal, 912, 98, \dodoi{10.3847/1538-4357/abee11}

\bibitem[{Foreman-Mackey(2021)}]{foreman-mackey_tinygp_2021}
Foreman-Mackey, D. 2021, tinygp — tinygp.
\newblock \url{https://tinygp.readthedocs.io/en/stable/}

\bibitem[{Foreman-Mackey {et~al.}(2017)Foreman-Mackey, Agol, Ambikasaran, \&
  Angus}]{foreman-mackey_fast_2017}
Foreman-Mackey, D., Agol, E., Ambikasaran, S., \& Angus, R. 2017, The
  Astronomical Journal, 154, 220, \dodoi{10.3847/1538-3881/aa9332}

\bibitem[{Freedman(2021)}]{freedman_measurements_2021}
Freedman, W.~L. 2021, The Astrophysical Journal, 919, 16,
  \dodoi{10.3847/1538-4357/ac0e95}

\bibitem[{Fryer \& Kalogera(2001)}]{fryer_theoretical_2001}
Fryer, C.~L., \& Kalogera, V. 2001, The Astrophysical Journal, 554, 548,
  \dodoi{10.1086/321359}

\bibitem[{Gair {et~al.}(2023)Gair, Ghosh, Gray, Holz, Mastrogiovanni,
  Mukherjee, Palmese, Tamanini, Baker, Beirnaert, Bilicki, Chen, Dálya,
  Ezquiaga, Farr, Fishbach, Garcia-Bellido, Ghosh, Huang, Karathanasis, Leyde,
  Magaña~Hernandez, Noller, Pierra, Raffai, Romano, Seglar-Arroyo, Steer,
  Turski, Vaccaro, \& Vallejo-Peña}]{gair_hitchhikers_2023}
Gair, J.~R., Ghosh, A., Gray, R., {et~al.} 2023, The Astronomical Journal, 166,
  22, \dodoi{10.3847/1538-3881/acca78}

\bibitem[{Giacobbo {et~al.}(2018)Giacobbo, Mapelli, \&
  Spera}]{giacobbo_merging_2018}
Giacobbo, N., Mapelli, M., \& Spera, M. 2018, Monthly Notices of the Royal
  Astronomical Society, 474, 2959, \dodoi{10.1093/mnras/stx2933}

\bibitem[{Gray {et~al.}(2022)Gray, Messenger, \& Veitch}]{gray_pixelated_2022}
Gray, R., Messenger, C., \& Veitch, J. 2022, Monthly Notices of the Royal
  Astronomical Society, 512, 1127, \dodoi{10.1093/mnras/stac366}

\bibitem[{Gray {et~al.}(2020)Gray, Hernandez, Qi, Sur, Brady, Chen, Farr,
  Fishbach, Gair, Ghosh, Holz, Mastrogiovanni, Messenger, Steer, \&
  Veitch}]{gray_cosmological_2020}
Gray, R., Hernandez, I.~M., Qi, H., {et~al.} 2020, Physical Review D, 101,
  122001, \dodoi{10.1103/PhysRevD.101.122001}

\bibitem[{Gray {et~al.}(2023)Gray, Beirnaert, Karathanasis, Revenu, Turski,
  Chen, Baker, Vallejo, Romano, Ghosh, Ghosh, Leyde, Mastrogiovanni, \&
  More}]{gray_joint_2023}
Gray, R., Beirnaert, F., Karathanasis, C., {et~al.} 2023, Journal of Cosmology
  and Astroparticle Physics, 2023, 023, \dodoi{10.1088/1475-7516/2023/12/023}

\bibitem[{Handcock \& Stein(1993)}]{handcock_bayesian_1993}
Handcock, M.~S., \& Stein, M.~L. 1993, Technometrics, 35, 403,
  \dodoi{10.2307/1270273}

\bibitem[{Hoffman \& Gelman(2011)}]{hoffman_no-u-turn_2011}
Hoffman, M.~D., \& Gelman, A. 2011, The {No}-{U}-{Turn} {Sampler}: {Adaptively}
  {Setting} {Path} {Lengths} in {Hamiltonian} {Monte} {Carlo},  arXiv.
\newblock \url{http://arxiv.org/abs/1111.4246}

\bibitem[{Holz \& Hughes(2005)}]{holz_using_2005}
Holz, D.~E., \& Hughes, S.~A. 2005, The Astrophysical Journal, 629, 15,
  \dodoi{10.1086/431341}

\bibitem[{Huerta {et~al.}(2018)Huerta, Moore, Kumar, George, Chua, Haas,
  Wessel, Johnson, Glennon, Rebei, Holgado, Gair, \&
  Pfeiffer}]{huerta_eccentric_2018}
Huerta, E.~A., Moore, C.~J., Kumar, P., {et~al.} 2018, Physical Review D, 97,
  024031, \dodoi{10.1103/PhysRevD.97.024031}

\bibitem[{Kiendrebeogo {et~al.}(2023)Kiendrebeogo, Farah, Foley, Gray, Kunert,
  Puecher, Toivonen, VandenBerg, Anand, Ahumada, Karambelkar, Coughlin,
  Dietrich, Kam, Pang, Singer, \& Sravan}]{kiendrebeogo_updated_2023}
Kiendrebeogo, R.~W., Farah, A.~M., Foley, E.~M., {et~al.} 2023, Updated
  observing scenarios and multi-messenger implications for the {International}
  {Gravitational}-wave {Network}'s {O4} and {O5},  arXiv,
  \dodoi{10.48550/arXiv.2306.09234}

\bibitem[{Kumar {et~al.}(2019)Kumar, Carroll, Hartikainen, \&
  Martin}]{kumar_arviz_2019}
Kumar, R., Carroll, C., Hartikainen, A., \& Martin, O. 2019, Journal of Open
  Source Software, 4, 1143, \dodoi{10.21105/joss.01143}

\bibitem[{Landry \& Essick(2019)}]{landry_nonparametric_2019}
Landry, P., \& Essick, R. 2019, Physical Review D, 99, 084049,
  \dodoi{10.1103/PhysRevD.99.084049}

\bibitem[{Li {et~al.}(2021)Li, Wang, Han, Tang, Yuan, Fan, \&
  Wei}]{li_flexible_2021}
Li, Y.-J., Wang, Y.-Z., Han, M.-Z., {et~al.} 2021, The Astrophysical Journal,
  917, 33, \dodoi{10.3847/1538-4357/ac0971}

\bibitem[{Loredo(2009)}]{loredo_handling_2009}
Loredo, T. 2009, 213, 211.04.
\newblock \url{https://ui.adsabs.harvard.edu/abs/2009AAS...21321104L}

\bibitem[{{Luger} {et~al.}(2021){Luger}, {Bedell}, {Foreman-Mackey},
  {Crossfield}, {Zhao}, \& {Hogg}}]{Luger2021}
{Luger}, R., {Bedell}, M., {Foreman-Mackey}, D., {et~al.} 2021, arXiv e-prints,
  arXiv:2110.06271.
\newblock \doarXiv{2110.06271}

\bibitem[{Magaña~Hernandez \& Ray(2024)}]{magana_hernandez_beyond_2024}
Magaña~Hernandez, I., \& Ray, A. 2024, Beyond {Gaps} and {Bumps}: {Spectral}
  {Siren} {Cosmology} with {Non}-{Parametric} {Population} {Models}.
\newblock \url{https://ui.adsabs.harvard.edu/abs/2024arXiv240402522M}

\bibitem[{Mandel {et~al.}(2017)Mandel, Farr, Colonna, Stevenson, Tiňo, \&
  Veitch}]{mandel_model-independent_2017}
Mandel, I., Farr, W.~M., Colonna, A., {et~al.} 2017, Monthly Notices of the
  Royal Astronomical Society, 465, 3254, \dodoi{10.1093/mnras/stw2883}

\bibitem[{Mandel {et~al.}(2019)Mandel, Farr, \& Gair}]{mandel_extracting_2019}
Mandel, I., Farr, W.~M., \& Gair, J.~R. 2019, Monthly Notices of the Royal
  Astronomical Society, 486, 1086, \dodoi{10.1093/mnras/stz896}

\bibitem[{Mapelli(2020)}]{mapelli_binary_2020}
Mapelli, M. 2020, Frontiers in Astronomy and Space Sciences, 7, 38,
  \dodoi{10.3389/fspas.2020.00038}

\bibitem[{Mapelli \& Giacobbo(2018)}]{mapelli_cosmic_2018}
Mapelli, M., \& Giacobbo, N. 2018, Monthly Notices of the Royal Astronomical
  Society, 479, 4391, \dodoi{10.1093/mnras/sty1613}

\bibitem[{Marchant \& Bodensteiner(2023)}]{marchant_evolution_2023}
Marchant, P., \& Bodensteiner, J. 2023, The {Evolution} of {Massive} {Binary}
  {Stars},  arXiv.
\newblock \url{http://arxiv.org/abs/2311.01865}

\bibitem[{Mastrogiovanni {et~al.}(2021)Mastrogiovanni, Leyde, Karathanasis,
  Chassande-Mottin, Steer, Gair, Ghosh, Gray, Mukherjee, \&
  Rinaldi}]{mastrogiovanni_importance_2021}
Mastrogiovanni, S., Leyde, K., Karathanasis, C., {et~al.} 2021, Physical Review
  D, 104, 062009, \dodoi{10.1103/PhysRevD.104.062009}

\bibitem[{Mastrogiovanni {et~al.}(2023{\natexlab{a}})Mastrogiovanni, Laghi,
  Gray, Santoro, Ghosh, Karathanasis, Leyde, Steer, Perriès, \&
  Pierra}]{mastrogiovanni_joint_2023}
Mastrogiovanni, S., Laghi, D., Gray, R., {et~al.} 2023{\natexlab{a}}, Physical
  Review D, 108, 042002, \dodoi{10.1103/PhysRevD.108.042002}

\bibitem[{Mastrogiovanni {et~al.}(2023{\natexlab{b}})Mastrogiovanni, Pierra,
  Perriès, Laghi, Santoro, Ghosh, Gray, Karathanasis, \&
  Leyde}]{mastrogiovanni_icarogw_2023}
Mastrogiovanni, S., Pierra, G., Perriès, S., {et~al.} 2023{\natexlab{b}},
  {ICAROGW}: {A} python package for inference of astrophysical population
  properties of noisy, heterogeneous and incomplete observations,  arXiv,
  \dodoi{10.48550/arXiv.2305.17973}

\bibitem[{Messenger \& Read(2012)}]{messenger_measuring_2012}
Messenger, C., \& Read, J. 2012, Physical Review Letters, 108, 091101,
  \dodoi{10.1103/PhysRevLett.108.091101}

\bibitem[{Mukherjee(2022)}]{Mukherjee:2021rtw}
Mukherjee, S. 2022, Mon. Not. Roy. Astron. Soc., 515, 5495,
  \dodoi{10.1093/mnras/stac2152}

\bibitem[{Neijssel {et~al.}(2019)Neijssel, Vigna-Gómez, Stevenson, Barrett,
  Gaebel, Broekgaarden, de Mink, Szécsi, Vinciguerra, \&
  Mandel}]{neijssel_effect_2019}
Neijssel, C.~J., Vigna-Gómez, A., Stevenson, S., {et~al.} 2019, Monthly
  Notices of the Royal Astronomical Society, 490, 3740,
  \dodoi{10.1093/mnras/stz2840}

\bibitem[{Perlmutter \& Schmidt(2003)}]{perlmutter_measuring_2003}
Perlmutter, S., \& Schmidt, B.~P. 2003, Supernovae and Gamma-Ray Bursters, 598,
  195, \dodoi{10.1007/3-540-45863-8_11}

\bibitem[{Phan {et~al.}(2019)Phan, Pradhan, \&
  Jankowiak}]{phan_composable_2019}
Phan, D., Pradhan, N., \& Jankowiak, M. 2019, Composable {Effects} for
  {Flexible} and {Accelerated} {Probabilistic} {Programming} in {NumPyro},
  \dodoi{10.48550/arXiv.1912.11554}

\bibitem[{Pierra {et~al.}(2023)Pierra, Mastrogiovanni, Perriès, \&
  Mapelli}]{pierra_study_2023}
Pierra, G., Mastrogiovanni, S., Perriès, S., \& Mapelli, M. 2023, A {Study} of
  {Systematics} on the {Cosmological} {Inference} of the {Hubble} {Constant}
  from {Gravitational} {Wave} {Standard} {Sirens},  arXiv,
  \dodoi{10.48550/arXiv.2312.11627}

\bibitem[{{Planck Collaboration} {et~al.}(2016){Planck Collaboration}, Ade,
  Aghanim, Arnaud, Ashdown, Aumont,
  {et~al.}}]{planck_collaboration_planck_2016}
{Planck Collaboration}, Ade, P. A.~R., Aghanim, N., {et~al.} 2016, Astronomy
  and Astrophysics, 594, A13, \dodoi{10.1051/0004-6361/201525830}

\bibitem[{Rasmussen \& Williams(2006)}]{rasmussen_gaussian_2006}
Rasmussen, C.~E., \& Williams, C. K.~I. 2006, Gaussian processes for machine
  learning, Adaptive computation and machine learning (Cambridge, Mass: MIT
  Press)

\bibitem[{Ray {et~al.}(2023)Ray, Magaña~Hernandez, Mohite, Creighton, \&
  Kapadia}]{ray_non-parametric_2023}
Ray, A., Magaña~Hernandez, I., Mohite, S., Creighton, J., \& Kapadia, S. 2023,
  Non-parametric inference of the population of compact binaries from
  gravitational wave observations using binned {Gaussian} processes, Tech.
  rep., \dodoi{10.48550/arXiv.2304.08046}

\bibitem[{Rinaldi \& Del~Pozzo(2022)}]{rinaldi_hdpgmm_2022}
Rinaldi, S., \& Del~Pozzo, W. 2022, Monthly Notices of the Royal Astronomical
  Society, 509, 5454, \dodoi{10.1093/mnras/stab3224}

\bibitem[{Sadiq {et~al.}(2022)Sadiq, Dent, \& Wysocki}]{sadiq_flexible_2022}
Sadiq, J., Dent, T., \& Wysocki, D. 2022, Physical Review D, 105, 123014,
  \dodoi{10.1103/PhysRevD.105.123014}

\bibitem[{Schutz(1986)}]{schutz_determining_1986}
Schutz, B.~F. 1986, Nature, 323, 310, \dodoi{10.1038/323310a0}

\bibitem[{Simpson(2022)}]{simpson_garcpas_2022}
Simpson, D. 2022, Un garçon pas comme les autres ({Bayes}) - {Priors} part 4:
  {Specifying} priors that appropriately penalise complexity.
\newblock
  \url{https://dansblog.netlify.app/posts/2022-08-29-priors4/priors4.html#the-dream-pc-priors-in-practice}

\bibitem[{Simpson {et~al.}(2017)Simpson, Rue, Riebler, Martins, \&
  Sørbye}]{simpson_penalising_2017}
Simpson, D., Rue, H., Riebler, A., Martins, T.~G., \& Sørbye, S.~H. 2017,
  Statistical Science, 32, 1, \dodoi{10.1214/16-STS576}

\bibitem[{Soares-Santos {et~al.}(2019)Soares-Santos, Palmese, Hartley, Annis,
  Garcia-Bellido, Lahav, Doctor, Fishbach, Holz, Lin, Pereira, Garcia, Herner,
  Kessler, Peiris, Sako, Allam, Brout, Rosell, Chen, Conselice,
  {et~al.}}]{soares-santos_first_2019}
Soares-Santos, M., Palmese, A., Hartley, W., {et~al.} 2019, The Astrophysical
  Journal Letters, 876, L7, \dodoi{10.3847/2041-8213/ab14f1}

\bibitem[{Spergel {et~al.}(2015)Spergel, Gehrels, Baltay, Bennett,
  Breckinridge, Donahue, Dressler, Gaudi, Greene, Guyon, Hirata, Kalirai,
  Kasdin, Macintosh, Moos, Perlmutter, Postman, Rauscher, Rhodes, Wang,
  Weinberg, Benford, Hudson, Jeong, Mellier, Traub, Yamada, Capak, Colbert,
  Masters, Penny, Savransky, Stern, Zimmerman, Barry, Bartusek, Carpenter,
  Cheng, Content, Dekens, Demers, Grady, Jackson, Kuan, Kruk, Melton, Nemati,
  Parvin, Poberezhskiy, Peddie, Ruffa, Wallace, Whipple, Wollack, \&
  Zhao}]{spergel_wide-field_2015}
Spergel, D., Gehrels, N., Baltay, C., {et~al.} 2015, Wide-{Field} {InfrarRed}
  {Survey} {Telescope}-{Astrophysics} {Focused} {Telescope} {Assets}
  {WFIRST}-{AFTA} 2015 {Report},  arXiv.
\newblock \url{http://arxiv.org/abs/1503.03757}

\bibitem[{Stein(1999)}]{stein_interpolation_1999}
Stein, M.~L. 1999, Interpolation of {Spatial} {Data}, Springer {Series} in
  {Statistics} (New York, NY: Springer), \dodoi{10.1007/978-1-4612-1494-6}

\bibitem[{Talbot \& Thrane(2018)}]{talbot_measuring_2018}
Talbot, C., \& Thrane, E. 2018, The Astrophysical Journal, 856, 173,
  \dodoi{10.3847/1538-4357/aab34c}

\bibitem[{{Tanvir} {et~al.}(2017){Tanvir}, {Levan},
  {Gonz{\'a}lez-Fern{\'a}ndez}, {Korobkin}, {Mandel},
  {et~al.}}]{2017ApJ...848L..27T}
{Tanvir}, N.~R., {Levan}, A.~J., {Gonz{\'a}lez-Fern{\'a}ndez}, C., {et~al.}
  2017, \apjl, 848, L27, \dodoi{10.3847/2041-8213/aa90b6}

\bibitem[{Taylor {et~al.}(2012)Taylor, Gair, \& Mandel}]{Taylor:2011fs}
Taylor, S.~R., Gair, J.~R., \& Mandel, I. 2012, Phys. Rev. D, 85, 023535,
  \dodoi{10.1103/PhysRevD.85.023535}

\bibitem[{Tiwari(2021)}]{tiwari_vamana_2021}
Tiwari, V. 2021, Classical and Quantum Gravity, 38, 155007,
  \dodoi{10.1088/1361-6382/ac0b54}

\bibitem[{{Torniamenti} {et~al.}(2024){Torniamenti}, {Mapelli}, {P{\'e}rigois},
  {Arca Sedda}, {Artale}, {Dall'Amico}, \& {Vaccaro}}]{2024arXiv240114837T}
{Torniamenti}, S., {Mapelli}, M., {P{\'e}rigois}, C., {et~al.} 2024, arXiv
  e-prints, arXiv:2401.14837, \dodoi{10.48550/arXiv.2401.14837}

\bibitem[{Valenti {et~al.}(2017)Valenti, Sand, Yang, Cappellaro, Tartaglia,
  Corsi, Jha, Reichart, Haislip, \& Kouprianov}]{valenti_discovery_2017}
Valenti, S., Sand, J., D., Yang, S., {et~al.} 2017, The Astrophysical Journal,
  848, L24, \dodoi{10.3847/2041-8213/aa8edf}

\bibitem[{van Son {et~al.}(2023)van Son, Mink, Chruślińska, Conroy, Pakmor,
  \& Hernquist}]{van_son_locations_2023}
van Son, L. A.~C., Mink, S. E.~d., Chruślińska, M., {et~al.} 2023, The
  Astrophysical Journal, 948, 105, \dodoi{10.3847/1538-4357/acbf51}

\bibitem[{van Son {et~al.}(2022)van Son, de~Mink, Callister, Justham, Renzo,
  Wagg, Broekgaarden, Kummer, Pakmor, \& Mandel}]{van_son_redshift_2022}
van Son, L. A.~C., de~Mink, S.~E., Callister, T., {et~al.} 2022, The
  Astrophysical Journal, 931, 17, \dodoi{10.3847/1538-4357/ac64a3}

\bibitem[{Vitale {et~al.}(2020)Vitale, Gerosa, Farr, \&
  Taylor}]{vitale_inferring_2020}
Vitale, S., Gerosa, D., Farr, W.~M., \& Taylor, S.~R. 2020, Inferring the
  properties of a population of compact binaries in presence of selection
  effects, \dodoi{10.1007/978-981-15-4702-7_45-1}

\bibitem[{{Ye} \& {Fishbach}(2024)}]{2024arXiv240212444Y}
{Ye}, C.~S., \& {Fishbach}, M. 2024, arXiv e-prints, arXiv:2402.12444,
  \dodoi{10.48550/arXiv.2402.12444}

\bibitem[{You {et~al.}(2021)You, Zhu, Ashton, Thrane, \& Zhu}]{You:2020wju}
You, Z.-Q., Zhu, X.-J., Ashton, G., Thrane, E., \& Zhu, Z.-H. 2021, Astrophys.
  J., 908, 215, \dodoi{10.3847/1538-4357/abd4d4}

\bibitem[{Zevin {et~al.}(2017)Zevin, Pankow, Rodriguez, Sampson, Chase,
  Kalogera, \& Rasio}]{zevin_constraining_2017}
Zevin, M., Pankow, C., Rodriguez, C.~L., {et~al.} 2017, The Astrophysical
  Journal, 846, 82, \dodoi{10.3847/1538-4357/aa8408}

\bibitem[{{Zevin} {et~al.}(2021){Zevin}, {Bavera}, {Berry}, {Kalogera},
  {Fragos}, {Marchant}, {Rodriguez}, {Antonini}, {Holz}, \&
  {Pankow}}]{2021ApJ...910..152Z}
{Zevin}, M., {Bavera}, S.~S., {Berry}, C. P.~L., {et~al.} 2021, \apj, 910, 152,
  \dodoi{10.3847/1538-4357/abe40e}

\bibitem[{Özel {et~al.}(2010)Özel, Psaltis, Narayan, \&
  McClintock}]{ozel_black_2010}
Özel, F., Psaltis, D., Narayan, R., \& McClintock, J.~E. 2010, 725, 1918,
  \dodoi{10.1088/0004-637X/725/2/1918}

\end{thebibliography}
\appendix
\section{Details of data simulation}
\label{ap:data generation}
The exact form of the injected population is
\begin{equation}
\label{eq:underlying population}
    \frac{\diff N}{\diff m_1 \diff z} \propto p(m_1|\bar{\Lambda}_m) p(z|\bar{\Lambda}_z, H_0, \Omega_M),
\end{equation}
where 
\begin{align}
    p(m_1,m_2|\bar{\Lambda}_m) &\propto \mathcal{S}(m_{\min},m_{\max})
    \left( f_{\text{peak}}e^{-\frac{1}{2}(\frac{m_1-\mu}{\sigma})^2}\mathcal{N}_{\text{g}} +
    (1-f_{\text{peak}})m_1^{\alpha}\mathcal{N}_{\text{pl}} \right) ,
\label{eq:underlying mass dist}
\end{align}
\begin{equation}
    p(z|H_0, \Omega_M) \propto \frac{\diff V_C}{\diff z} \frac{1}{1+z} \frac{(1+z)^\alpha_z}{1+\left(\frac{1+z}{1+z_p}\right)^{\alpha_z+\beta_z}}.
    \label{eq:underlying redshift dist}
\end{equation}
Here, $V_C(H_0, \Omega_M)$ is the comoving volume for given cosmological parameters \Ho{} and \Omm{}, and $\bar{\Lambda}_m = \{\alpha, m_{\min}, m_{\max}, \mu, \sigma, f_{\text{peak}}\}$ are the (hyper-)parameters describing the power law in primary mass, minimum and maximum black hole mass, Gaussian peak location and width, and fraction of events in the Gaussian peak, respectively.
$\mathcal{S}$ is a smoothing function at low and high masses, and $\mathcal{N}_{\text{pl}}$ and $\mathcal{N}_{\text{g}}$ are the normalizations between  $m_{\min}$ and $m_{\max}$ for the power law component and truncated Gaussian component, respectively.
The smoothing creates support for masses below $m_{\min}$ and above $m_{\max}$.
$\bar{\Lambda}_z = \{z_p,\alpha_z,\beta_z\}$ are the parameters governing the peak of the redshift distribution, low-$z$ power law slope, and high-$z$ power law slope.
When generating the simulated events, we have fixed $\alpha=-2.7$, $m_{\max}=78\Msun$, $m_{\min}=10~\Msun$, $\mu=30~\Msun$, $\sigma=7.0~\Msun$, $f_{\text{peak}}=0.05$, $z_p=2.4$, $\alpha_z=1.0$, and $\beta_z=3.4$.
These choices correspond to the maximum \emph{a posteriori} values obtained by an analysis of GWTC-3 data using the \plp{} model \citep{abbott_population_2023}.
For simplicity, we assume \new{a uniform mass ratio distribution.}
Thus, $\theta = \{m_1,z\}$.
We consistently apply these assumptions to the data generation process and the population inference.
We do not fit for or simulate spins, as they do not redshift and hence do not carry additional cosmological information, and \new{we fix the distribution of mass ratios.}

We use cosmological parameters \Ho$=%
  67.66
\unskip\label{output/H0_FID.txt}\unskip%
\Hunits$, \Omm$=0.3$, and $\Omega_\Lambda=1-$\Omm, consistent with those found by \citet{planck_collaboration_planck_2016}.
We emphasize, however, that the choice of cosmological parameters for data generation is arbitrary and does not impact the results, since we are concerned only with the ability of our method to recover the injected values.
Throughout the data generation and inference, we use the approximations presented in \citet{adachi_analytical_2012} to efficiently convert between $D_L$ and $z$ for a given set of cosmological parameters \Omm{} and \Ho.

We do not include neutron stars in our simulation set, as their contribution to the spectral siren measurement is expected to be subdominant in \ac{O5}.
However, if a lower mass gap between the heaviest neutron stars and lightest black holes exists, it will provide an additional feature with which to inform the measurement, and will be the most informative feature for spectral siren measurements with next-generation detectors \citep{ezquiaga_spectral_2022}.

After passing the simulated events through projected detector selection effects, the resulting catalog has $%
  591
\unskip\label{output/num_found_events.txt}\unskip%
$ events, consistent with the numbers projected for \ac{O5} by \citet{kiendrebeogo_updated_2023}.
We use the software package \texttt{GWMockCat} \citep{farah_things_2023} to simulate posterior samples for these events with measurement uncertainties typical of those expected from \ac{O5} detectors.
\texttt{GWMockCat} also simulates a set of software injections, which we use to estimate selection effects in the inference.
To determine the detectability of both injections and simulated events in O5, we use the projected \ac{O5} LIGO power spectral density \citep{obsscen_noise_curves,abbott_prospects_2020} for a single detector to calculate observed signal-to-noise ratios $\rho_{\text{obs}}$, and we consider events and injections with a single-detector SNR $\rho_{\text{obs}}>8$ to be detectable. 
The full procedure for this mock data generation process is described in \citet{fishbach_where_2017, farah_things_2023, essick_ensuring_2024}.

\section{Effects of fitting for secondary mass}
\label{ap:mass ratio}
\new{We examine the effect of only fitting the distribution of primary masses on our results.
To do so, we perform two spectral siren analyses: one that includes a fit to mass ratio, and one that does not.
Both analyses use the parametric, \plp{} model for the distribution of primary masses.
We model the distribution of mass ratios with a power law, similarly to the majority of analyses presented in \citet{abbott_population_2021,abbott_population_2023}.}

\begin{figure*}
    \includegraphics[width=\textwidth]{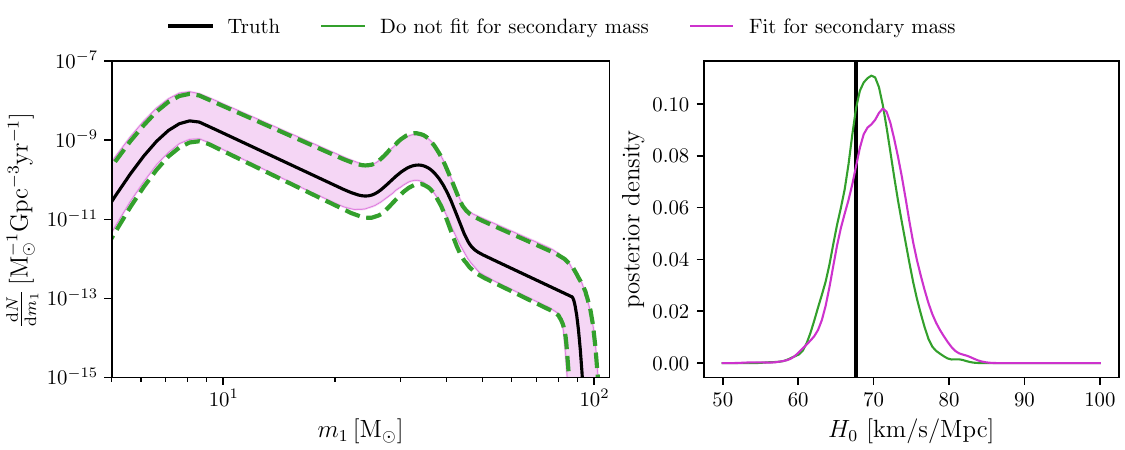}
    \caption{Comparison of a parametric spectral siren analysis performed while fitting for the distributions of primary and secondary masses (magenta solid) and primary masses only (green dashed).
    The posterior on \Ho{} is relatively unchanged between the two, and only minor differences are observed in the mass spectra.}
    \label{fig:mass ratio}
    \script{fit_q_comparison.py}
\end{figure*}

\new{We use the same set of simulated events for both analyses.
These are generated in the same way as described in Section~\ref{sec:data generation}.
The inference results are shown in Figure~\ref{fig:mass ratio}.
We find the posteriors on \Ho{} to be similar between the two cases.
Additionally, the recovered mass spectra are nearly identical.
We therefore conclude that fitting for mass ratios does not significantly impact our main conclusions.
}

\section{Gaussian process-based mass distribution}
\label{ap:GP}

In this Section, we discuss the properties of the \ac{GP}-based mass distribution and describe our modeling choices in more detail. 

Practically, the difference in the inference of the population when using a \ac{GP} versus other modeling choices is that the population model ($\diff N/\diff t_{\det} \diff \theta_i (\theta_i;\Lambda)$ in Equation~\ref{eq:inhomog-poisson}) is determined directly by a realization of the \ac{GP}, rather than by a handful of hyper-parameters $\Lambda$ and evaluated on an analytical function.
In other words, when using parametric models, $\diff N/\diff t_{\det} \diff \theta_i (\theta_i;\Lambda)$ is calculated by evaluating a specific functional form described by a small set of hyper-parameters. 
With the \ac{GP} approach, the hyper-parameters describing the mass distribution are the rate at each event-level posterior sample's source frame mass, and the rate at each found injection's source frame mass.
The \ac{GP} \emph{is} $p(\Lambda)$, the prior on population parameters (except in the case of cosmological parameters, which all have uniform priors).

Because the \ac{GP} is defined only at specific data points, we have $N_{\text{ev}}M + N_{\text{inj}}$ mass hyper-parameters, where $M$ is the number of posterior samples per event and $N_{\text{inj}}$ is the number of injections used to calculate the selection function~\citep[see e.g.][]{vitale_inferring_2020,essick_estimating_2021}.
In this way, our \ac{GP}-based mass distribution is similar to the autoregressive population models used in~\citet{callister_parameter-free_2023}.
Indeed, an autoregressive process is a \ac{GP} with a specific choice of kernel.

The kernel is a function that defines the covariance between input points in the \ac{GP} (in our case, two source frame mass values). 
It defines the notion of similarity between adjacent points and thereby encodes our assumptions about the smoothness of the source frame mass distribution \citep{rasmussen_gaussian_2006}.
We use a Mat\'ern kernel \citep{handcock_bayesian_1993, stein_interpolation_1999} with $\nu = 5/2$, but have repeated the analysis with $\nu=3/2$ and $\infty$, finding little impact on the results, except that the $\nu=\infty$ case (also called the squared exponential kernel) produces a slightly more jagged mass distribution.
In addition to the mean, Mat\'ern kernels have two parameters that determine their properties: a length scale $l$ and a variance $s$.
In our use case, these are one level further removed from hyper-parameters, so we adopt the terminology used in \citet{callister_parameter-free_2023} and call them ``hyper-hyper-parameters.''
We fit these hyper-hyper-parameters along with the hyper-parameters $\Lambda$ to minimize prior assumptions about the form of the mass distribution.
We use penalized-complexity priors on the hyper-hyper-parameters to enforce that the model does not create small-scale structure uninformed by data, thereby avoiding over-fitting \citep{simpson_penalising_2017, simpson_garcpas_2022}. 
Explicitly, the priors on $l$ and $s$ are Fr\'echet and Gamma distributions, respectively, and are defined to have less than 5\% support for correlation lengths smaller than the average spacing between event-level posterior means.

The time to evaluate a \ac{GP} is notorious for scaling as the cube of the number of data points, making \acp{GP} unwieldy with large data sets, such as the $\mathcal{O}(10^9)$ posterior samples and software injections expected for \ac{O5}.
We therefore make two approximations to a full \ac{GP} to increase computational efficiency.
First, for each likelihood evaluation, we evaluate a full \ac{GP} on a regular grid between $0.1\Msun$ and $250\Msun$ and then interpolate it at each data point.
Second, we use the quasi-separability of Mat\'ern kernels to analytically perform the transformation between covariance matrix and \ac{GP} draw.
This second step is done using the \texttt{QuasisepSolver} module \citep{foreman-mackey_fast_2017} in the \texttt{tinygp} code base \citep{foreman-mackey_tinygp_2021}, and requires data to be sortable (i.e. one-dimensional).

Algorithmically, each posterior evaluation contains the following steps: %
   
\unskip\label{output/priors_placeholder.txt}\unskip%

\begin{enumerate}
\script{nonparametric_inference.py}
    \item Draw cosmological parameters \Ho{} and \Omm{} from uniform prior distributions.
    \item Convert the luminosity distances and detector-frame masses of each event posterior sample to redshifts and source-frame masses according to the cosmology specified by step 1.
    \item Draw hyper-hyper-parameters $l$ and $s$ from the penalized-complexity priors described above.
    \item Draw a single \ac{GP} realization with a kernel defined by $l$ and $s$.
    This is defined on a regular grid of source-frame masses and evaluated using the \texttt{QuasisepSolver} in \texttt{tinygp}.
    \item Interpolate the \ac{GP} at each event posterior sample and injection source-frame mass (from step 3).
    \item Calculate the population likelihood according to Equation~\ref{eq:inhomog-poisson}.
\end{enumerate}
We perform these steps within \texttt{numpyro} \citep{bingham_pyro_2019,phan_composable_2019}, sampling the posterior using the no-u-turn sampler for Hamiltonian Monte Carlo \citep{hoffman_no-u-turn_2011}. 
This can be seen explicitly in the source code accompanying this paper, in the \texttt{scripts/priors.py} %
  \unskip\label{output/priors_placeholder.txt}\unskip%
script.

\section{Biases induced by the \bpl{} model}
\label{ap:parametric bias}
\begin{figure}
    \centering
    \includegraphics[width=0.5\linewidth]{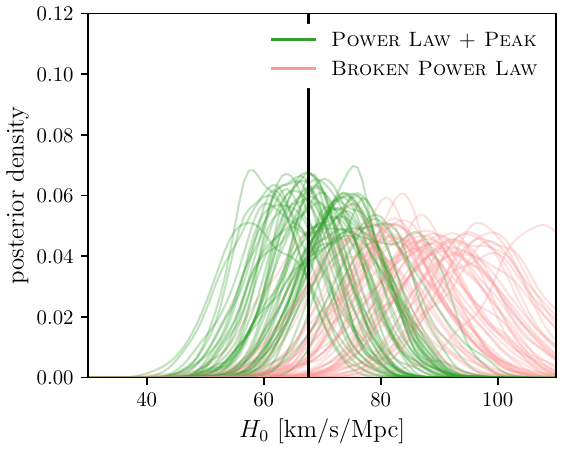}
    \caption{Recovered posteriors on \Ho{} from spectral siren analyses on 50 mock catalogs with the \plp{} (green) and \bpl{} (pink) parametric models. The true, injected value is shown by a vertical black line.}
    \label{fig:parametric bias}
    \script{bias_numbers.py}
\end{figure}

\new{Figure~\ref{fig:parametric bias} shows recovered posteriors on \Ho{} as inferred from 50 mock catalogs using both the \bpl{} and \plp{} parametric models.
We find that the \plp{} model recovers \Ho{} posteriors that are largely symmetric about the true value, whereas the \bpl{} model typically finds more support for larger \Ho{} values.
We note that \citep{pierra_study_2023} report an underestimate of \Ho{} when fitting their Broken Power Law model to a single dataset generated with \plp, but the prior bounds on the maximum mass parameter of their model do not extend below the injected value of that parameter, which in turn does 
not allow for the possibility of an overestimated \Ho{}.
It is therefore not possible to directly compare the direction of our observed \Ho{} offset with theirs, despite the similarities bewteen our chosen population models.}

\new{Regardless, this systematic offset demonstrates that incorrect parametric models unfortunately induce a systematic bias in cosmological inference, not an increase in statistical uncertainty. 
Other parametric models may show different trends depending on which parts of the mass distribution they incorrectly model and how.
Therefore, while we are able to identify the shortcomings of \bpl{} when \plp{} describes the true mass distribution, it is not possible in practice to know which direction an incorrect parametric model will bias \Ho{}.
Mitigating the bias by using more flexible models may thus be the simplest solution to this problem.}

\end{document}